\documentclass[useAMS,usenatbib]{mnras}
\usepackage{epsfig,latexsym, amssymb, amscd, psfrag,natbib}
\usepackage{longtable}

\def\kmsmpc{km\,s$^{-1}$\,Mpc$^{-1}$}
\def\eg{{e.g.,}}
\def\ie{{i.e.,}}
\def\etal{{et~al.}}
\def\hii{\ion{H}{ii}}
\def\m{{\,\micron}} 

\def\nuv{NUV }

\def\hab{H$\alpha$/H$\beta$}
\def\Msol{{\mbox{M$_\odot$}}}
\def\Lsol{{\mbox{L$_\odot$}}}
\def\Her{\mbox{\it Herschel}}
\def\Sp{\mbox{\it Spitzer}}
\def\IRAS{\mbox{\it IRAS}}
\def\WISE{\mbox{\it WISE}}
\def\PSCz{\mbox{PSC$z$}}
\def\lfir{\mbox{$L_{\rm FIR}$}}
\def\ltir{\mbox{$L_{\rm TIR}$}}
\def\lpah{\mbox{$L_{\rm PAH}$}}
\def\sfrr{\mbox{SFR$_{\rm 1.4\,GHz}$}}
\def\sfrrd{\mbox{SFR2$_{\rm 1.4\,GHz}$}}
\def\sfrp{\mbox{SFR$_{\rm PAH}$}}
\def\sfrt{\mbox{SFR$_{\rm tot}$}}
\def\sfrfir{\mbox{SFR$_{\rm FIR}$}}
\def\sfrfuv{\mbox{SFR$_{\rm FUV}$}}
\def\sfrnuv{\mbox{SFR$_{\rm NUV}$}}
\def\sfru{\mbox{SFR$_{U}$}}

\def\0{\phantom{$-$}}

\newcommand{\text}{}

\newcommand{\new}[1]{{#1}} 

\sfcode`A=1000 \sfcode`B=1000 \sfcode`C=1000 \sfcode`D=1000
\sfcode`E=1000 \sfcode`F=1000 \sfcode`G=1000 \sfcode`H=1000
\sfcode`I=1000 \sfcode`J=1000 \sfcode`K=1000 \sfcode`L=1000
\sfcode`M=1000 \sfcode`N=1000 \sfcode`O=1000 \sfcode`P=1000
\sfcode`Q=1000 \sfcode`R=1000 \sfcode`S=1000 \sfcode`T=1000
\sfcode`U=1000 \sfcode`V=1000 \sfcode`W=1000 \sfcode`X=1000
\sfcode`Y=1000 \sfcode`Z=1000

\title[Star Formation in Nearby Galaxies]
{The Star Formation Reference Survey III: A Multi-wavelength View of
 Star Formation in Nearby Galaxies}
\author[Mahajan et al.]
{\parbox{\textwidth}{Smriti Mahajan$^{1,2}$\thanks{E-mail: \texttt{smritimahajan@iisermohali.ac.in}}, M. L. N. Ashby$^{2}$, S. P. Willner$^{2}$, P. Barmby$^{3}$, 
 G. G. Fazio$^{2}$, A. Maragkoudakis$^{3,4,5}$, S. Raychaudhury$^{6}$, A. Zezas$^{4,5}$ 
 } \vspace{0.4cm} \\
\parbox{\textwidth}{$^{1}$Department of Physical Sciences, Indian Institute of Science Education and Research Mohali,
 Knowledge City, Sector 81, Manauli 140306, Punjab, India \\
 $^{2}$Harvard-Smithsonian Center for Astrophysics, 60 Garden Street,
 Cambridge, MA 02138 USA \\
$^{3}$Department of Physics and Astronomy, University of Western Ontario, London, Ontario, N6A 3K7 Canada \\ 
$^{4}$University of Crete, Department of Physics, GR-71003 Heraklion, Greece \\
$^{5}$Foundation for Research and Technology---Hellas (FORTH), Heraklion 71003, Greece \\
$^{6}$Inter-University Centre for Astronomy \& Astrophysics, Ganeshkhind, Pune, India }}

\date{Accepted 2018 September 11. Received 2018 August 31; in
  original form  2018 May 1}

\pubyear{2018}

\begin{document}
\label{firstpage}
\pagerange{\pageref{firstpage}--\pageref{lastpage}}
\maketitle


\begin{abstract}
We present multi-wavelength global star formation rate (SFR)
estimates for 326 galaxies from the Star Formation Reference Survey
(SFRS) {in order to determine the mutual scatter and range of
validity of different indicators.} The widely used empirical SFR
recipes based on 1.4\,GHz continuum, 8.0\,$\mu$m polycyclic
aromatic hydrocarbons (PAH), and a combination of far-infrared
(FIR) plus ultraviolet (UV) emission are {mutually consistent}
with scatter of $\la$0.3\,dex. The scatter is even smaller,
$\la$0.24\,dex, in the intermediate luminosity range
$9.3<\log(L_{60\,\micron}/\Lsol)<10.7$.  The data prefer a non-linear
relation between 1.4\,GHz luminosity and other SFR measures. PAH
luminosity underestimates SFR for galaxies with strong UV
emission. A bolometric extinction correction to far-ultraviolet
luminosity yields SFR within $0.2$\,dex of the total SFR estimate, but
extinction corrections based on UV spectral slope or nuclear Balmer
decrement give SFRs that may differ from the total SFR by up to
2\,dex. However, for the minority of galaxies with UV
luminosity ${>}5\times10^9$\,\Lsol\ or with implied far-UV
extinction $<$1\,mag, the UV spectral slope gives extinction
corrections with 0.22\,dex uncertainty.
\end{abstract}




\section{Introduction}
\label{intro}

Star formation is critical for galaxy evolution. Stars have created
almost all the elements heavier than helium in the Universe and play
a key role in recycling dust and metals in galaxies. Hence the rate
at which a galaxy forms stars is one of the most important drivers of
its evolution. Understanding global trends in star formation rate
(SFR henceforth) among different galaxy populations is required for
interpreting the `Hubble sequence,' which is a representation of
not just the evolutionary trend in galaxy morphology but also gas
content, mass, bars, dynamical structure, and environment, all of
which influence the SFR \citep[][and references
therein]{kennicutt98}. SFR measurements and the star formation rate
density are therefore essential for constraining the models of
structure formation in the Universe. However, until a few years ago,
accurate measurements of SFR even in nearby galaxies were difficult
owing to lack of knowledge of the effect of dust on different SFR
tracers.

In addition to the standard optical spectral lines (\eg~H$\alpha$,
[\ion{O}{ii}]), the indicators most commonly used to quantify star
formation in a galaxy are the global radio continuum, mid- and
far-infrared (MIR and FIR), and ultraviolet (UV) emission. Different
wavelengths trace stellar populations at different stages of
evolution as well as different galaxy components. For instance, stars
more massive than $\sim$8\,\Msol\ produce the \new{core-collapse}
supernovae whose remnants (SNRs) accelerate relativistic electrons,
which have lifetimes $\la$100\,Myr \citep{condon92}. The resulting
non-thermal radio synchrotron emission, which dominates a galaxy's
radio luminosity at low frequencies ($\la$5\,GHz), is therefore a
measure of past formation of massive stars. Thermal radio emission,
which dominates at high radio frequencies \citep[$\ga$10\,GHz;
\eg][]{klein81,gioia82,Tabatabaei2017} is a measure of current
production of ionizing photons. The massive stars producing such
photons have lifetimes of order 10\,Myr, and high-frequency radio
observations thus probe very recent SFR in star-forming and normal
galaxies.\footnote{In the context of this paper the term `normal' is
  used for galaxies without a strong active galactic nucleus (AGN)
  and with $0.1 \la {\rm SFR} \la 10$\,\Msol\ yr$^{-1}$.}

UV light is emitted predominantly by stars younger than around
$200$\,Myr and is therefore a good measure of SFR over time-scales of
tens of Myr. But the interpretation of this indicator is hampered by the
presence of dust clouds enshrouding young star-forming
regions. Dust absorbs UV photons and reemits their energy at FIR
wavelengths, making   FIR luminosity a
more reliable SFR indicator. For most star-forming galaxies, a
combination of UV and FIR luminosity accounts for a major fraction of
the galaxy's bolometric luminosity.

Generally the FIR emission from any galaxy has at least two
components, one originating from the interstellar dust heated by the
diffuse radiation field and a second contribution from star formation
activity in and near the \ion{H}{ii} regions \citep[][and references
therein]{soifer87}.  \new{If the second component can be measured,}
its FIR luminosity can be converted to an SFR measure.

Numerous attempts have been made to utilize the FIR--UV energy budget
to quantify dust attenuation in various samples of galaxies selected
at different wavelengths
\citep[\eg][]{xu95,meurer99,buat02,buat05,Cortese2006,dacunha10}. One
approach is 
to use the FIR/UV flux ratio \citep[or the infrared excess
IRX\footnote{In what follows ${\rm IRX} \equiv \log({\rm FIR/FUV})$
  {with FIR and FUV expressed in $\nu F_\nu$ units}.}
as it is more popularly known, \eg][]{meurer99,kong04,seibert05}. An
alternative is to use reddening inferred from the Balmer decrement
\citep[\eg][]{buat02,gilbank10} or from UV colour
\citep[\eg][]{meurer99,lee09,gilbank10}. The reddening measure is
then combined with an assumed extinction curve to yield extinction at
UV wavelengths.  The problem is understanding both random and
systematic errors for the derived extinction values for
different galaxy populations.

In order to understand the relation between indicators of star
formation and dust extinction, the different SFR and extinction
indicators need to be quantified and compared for a {\em{statistical
    sample}} of galaxies covering a {\em wide range} in physical and
intrinsic properties and having {\em{known biases}}. This idea
motivated the Star Formation Reference Survey \citep[SFRS;][Paper~I
  henceforth]{ashby11}. Which SFR indicators can be used to estimate the
global SFR of a galaxy? When are multi-wavelength data required? How
closely does any single SFR indicator measure a galaxy's `total' SFR?
Is the relation between individual SFR indicators universal
for all types of star-forming galaxies? What are the advantages and
disadvantages  of different extinction indicators?

\new{The primary purpose of}
this work is to present {\it GALEX} ultraviolet photometry for SFRS
galaxies.  By combining {\it GALEX} photometry with photometry at
other wavelengths, we test and mutually calibrate widely used
empirical formulas to calculate global SFRs for galaxies using
tracers spanning all available wavelengths. In choosing among the
many calibrations available in the literature, we have preferred
those that give mutually consistent results.  The
wide ranges of morphologies, luminosities, sizes, SFRs, and stellar
masses spanned by the SFRS galaxies, together with the sample's
well-defined selection criteria, makes it an ideal sample to quantify
the relation between different SFR measures in nearby galaxies and
hence a benchmark for comparing the SFR measures of high redshift
galaxies.  Studies such as this one
have been performed elsewhere
\citep[\eg][]{hopkins01,bell03,schmitt06,johnson07,zhu08,davies16,wang16}
but on samples often small or chosen without well-defined criteria or
with narrow sample boundaries, where systematic deviations from the
underlying correlations cannot be well explored.
{A recent study by \citet{Brown2017} used {\it GALEX} photometry
  as a SFR tracer, but their sample was restricted to galaxies with
  strong emission lines, and they did not use FIR at all.}

This paper is organized as follows. \S\ref{data}  describes
the datasets used in this paper, and \S\ref{s:sfr} describes
and compares the star formation tracers.  \S\ref{s:extinc}
investigates extinction indicators and whether they can give useful
measures of SFR. These are followed by a discussion of our findings
in the context of the existing literature in \S\ref{discussion},
followed by a summary of our results in \S\ref{conclusions}.
Throughout this paper, star formation rates are based on a Salpeter
IMF in the range 0.1--100\,\Msol.
 

\section{The data}
\label{data}

\subsection{Sample selection}
\label{ss}

The SFRS (Paper~I) is a statistically robust, representative sample
of 367 star-forming galaxies in the
local Universe. The sample selection criteria were defined
objectively to guarantee that the SFRS has {\em{known biases}}
and selection weights, making it possible to relate conclusions
  from the SFRS to magnitude-limited or volume-limited FIR-selected
  samples. Moreover, the SFRS spans the full ranges of properties
exhibited by FIR-selected star-forming galaxies in the nearby
Universe. While much larger galaxy samples exist, for huge
  samples it is difficult to obtain the complete data sets needed to
  explore multi-wavelength correlations. The SFRS is therefore an
ideal tool for understanding the global properties of nearby
($z\la0.1$) star-forming galaxies.

\begin{figure*}
\centering{{\rotatebox{0}{\epsfig{file=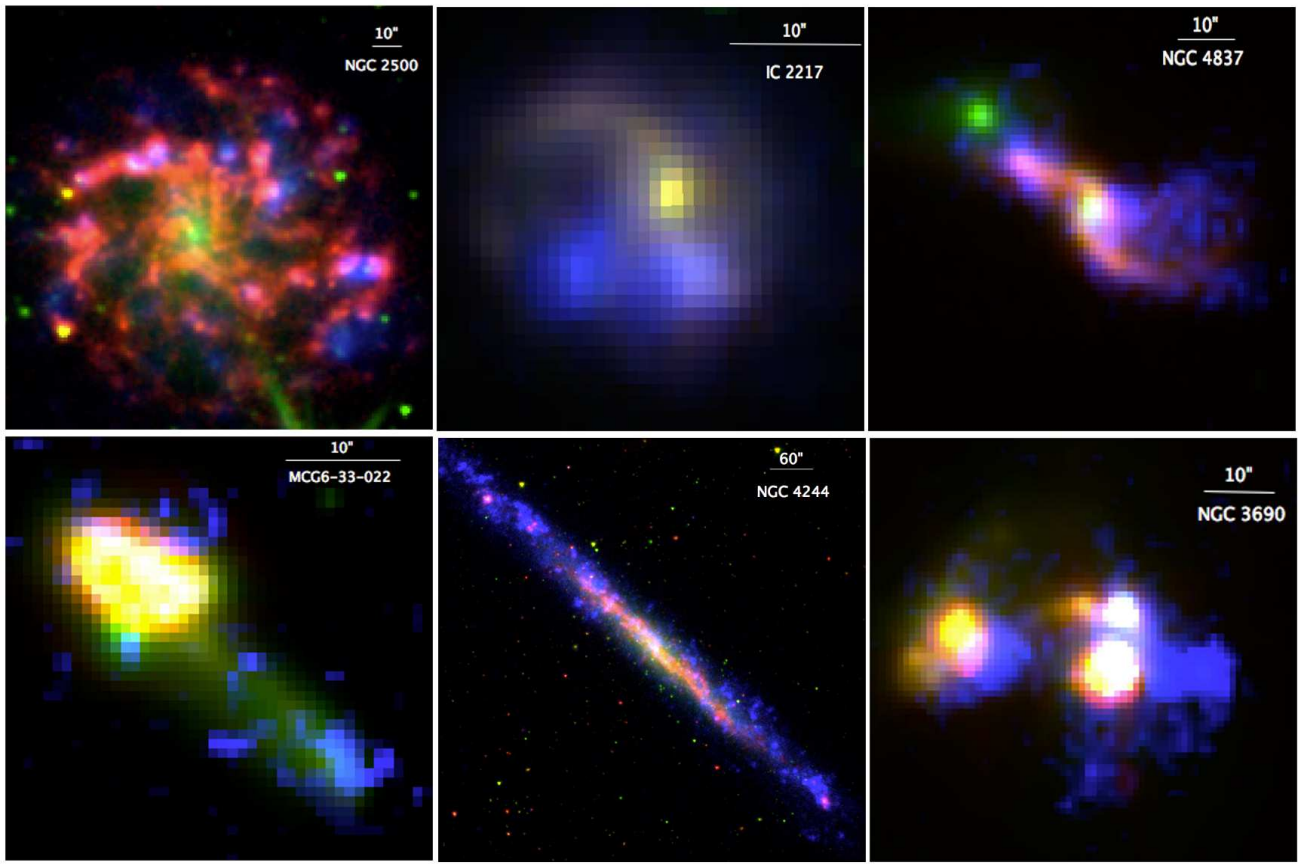,width=\textwidth}}}}
\caption{Six SFRS galaxies showing some of the variety in galaxy
  properties covered by this sample. colours represent IRAC 8.0\m\
  {\it{(red)}}, 3.6\m\ {\it{(green)}}, and {\it{GALEX}} FUV
  {\it{(blue)}}. Scale bars in each panel indicate angular sizes,
  which correspond to linear sizes of 0.72\,kpc for NGC~2500,
  3.7\,kpc for IC~2217, 6.4\,kpc for NGC~4837, 9.4\,kpc for
  MCG~6-33-022, 1.25\,kpc for NGC 4244, and 2.6\,kpc for NGC~3690. }
\label{image}
\end{figure*}
  
The SFRS was drawn from the \PSCz\ catalog \citep{saunders00}, an
all-sky redshift survey of 15,000 galaxies observed by \IRAS\ and
brighter than 0.6\,Jy at 60\m. From this was drawn a representative
subsample spanning the entire three-dimensional space formed by the
60\m\ luminosity $L_{60}$, flux ratio $F_{60}/K_s$, and the \IRAS\
flux density ratio $F_{100}/F_{60}$.  $L_{60}$ is a proxy for the
SFR, $F_{60}/K_s$ for specific star formation rate (sSFR), and
$F_{100}/F_{60}$ measures FIR colour temperature {($T_d$) and thus
  heating intensity within star formation regions
  (Paper~I)}. {$T_d$ increases with increasing far infrared
  luminosity \citep{Sanders1996}, and thus $T_d$} may be related to
the mode {(`normal' or `starburst' as described for example
  by \citealt{Daddi2010}, \citealt{Rodighiero2011}, or
  \citealt{Elbaz2011})} of star formation.  Although the {SFRS}
sample was drawn from FIR observations, it was constructed to include
representative numbers of galaxies with $L_{60} \la
10^{6.5}$\,\Lsol. Full details of the SFRS sample selection including
distance estimates are given in Paper~I.\footnote{The overall
  distance scale makes very little difference because we are dealing
  with ratios of SFR indicators. Paper~I distances were based on a
  variety of measures for nearby galaxies, where the Hubble distance
  is inaccurate, or on $H_0 = 73$\,km\,s$^{-1}$\,Mpc$^{-1}$ for more
  distant galaxies.} Figure~\ref{image} gives a glimpse of the wide
range in galaxy sizes and morphologies covered by the
SFRS. Metallicities (expressed as $\rm \log[O/H]+12$) range from
8.0 to 9.3 as measured by the `N2' method \citep{Marag2017b}.

Because the SFRS sample was constructed to span the known ranges of
galaxy SFR, sSFR, and $T_d$, uncommon types of galaxies
\new{(including, for example, edge-on galaxies) are over-represented
  in the sample. Galaxies with AGN are also included in the SFRS.  In
  fact, the quasar 3C~273 and blazar OJ~287 were selected by the SFRS
  criteria, where each occupies its own unique cell in the selection
  matrix. Because these objects are dominated by luminous AGNs, they
  are not relevant to studies of local star formation and are
  excluded from this paper (decreasing the sample size to 367
  galaxies). Other AGNs would not be as easily recognizable absent
  spectroscopic information.  For example,} SFRS 263 (=IRAS
13218+0552) is a Type-1 AGN \citep{Marag2017b}, and SFRS 270 (=IRAS
13349+2438) is a QSO \citep[e.g.,][]{Lee2013}.  These are retained in
our sample and will add to the observed scatter.

Paper~I gives weights for SFRS galaxies which, if applied, make the
sample proportional to the \citet{saunders00} \PSCz\
catalog. However, because the purpose of this paper is to test how
well empirical SFR metrics work for all types of star-forming
galaxies found in the local Universe, we have {\em not} applied these
weights for the numerical calculations. \new{This will make the
  derived scatter larger than would be the case for an unweighted
  sample, but the scatter will indicate the range that galaxies can
  occupy.}

\subsection{Ultraviolet data}

The ultraviolet images for SFRS galaxies were retrieved from data
releases 4/5 and 6 of the {\it{Galaxy Evolution EXplorer}}
\citep[{\it{GALEX}};][]{martin05,morrissey05}. {\it{GALEX}} conducted
an all-sky imaging survey along with targeted programs in two
photometric bands: 1516\,\AA\ (`far ultraviolet' or FUV) and
2267\,\AA\ (`near ultraviolet' or NUV). The bulk of the SFRS sample consists of bright, nearby
galaxies, and therefore no exposure time or brightness limit
constraints were imposed while looking for a {\it{GALEX}}
detection.  Almost three quarters of
the UV imaging data used in this paper were taken as part of the
{\it{GALEX}}'s primary All-Sky Imaging Survey (AIS) with an effective
exposure time of ${\sim}0.1$\,ks.  Most of the rest of the data come
from the Nearby Galaxies Survey (NGS) with an effective exposure time
of ${\sim}1.5$\,ks. The remainder were observed as a part of other
{\it{GALEX}} surveys as well as individual guest investigator
programs. We imposed no constraint on the location of galaxy in the
1\fdg2 {\it{GALEX}} field of view even though the point spread
function varies across {\it{GALEX}} images, thus requiring
non-negligible aperture corrections for faint sources detected away
from the image centre. Whenever possible, we chose images where the
target galaxy was closer to the centre of the field of view. Whenever
a galaxy was observed as a part of more than one program, we chose
the deepest observations. In total, {\it{GALEX}}
imaging data were available for 326/367 (89 per cent) of the sample galaxies in at least
one waveband.\footnote{NGC~3758 was undetected in NUV. In the FUV
  band, seven galaxies were observed but undetected, and two were not
  observed.} These 326 galaxies form the sample for this
paper. \new{The unobserved galaxies are mostly those near bright,
  blue stars that precluded {\it{GALEX}} imaging. Because this is a
  purely local effect, it should not bias our conclusions.}
Adopted distances and UV photometry for the sample galaxies are given
in Table~\ref{t:uv}.
 
\begin{table*}
\begin{center}
\caption{{\it GALEX} data for SFRS galaxies (complete table available
  in appendix).} \label{t:uv}
\begin{tabular}{clrccccc}
\hline\hline
SFRS$^1$ & 
Name & 
$D$ (Mpc)$^1$ &
FUV$^2$ & 
$\Delta$FUV & 
NUV$^2$ & 
$\Delta$NUV
& $E(B-V)^3$  \\
\hline
1&IC 486&114.4&18.179&0.029&17.575&0.014&0.040 \\
2&IC 2217&76.1&16.434&0.013&15.939&0.006&0.041 \\
3&NGC 2500&15.0&13.925&0.004&13.785&0.002&0.040 \\
5&MCG 6-18-009&164.4&17.890&0.026&17.064&0.011&0.052 \\
8&NGC 2532&77.6&15.417&0.008&14.862&0.004&0.054 \\
9&UGC 4261&93.2&16.485&0.014&16.159&0.007&0.055 \\
10&NGC 2535&61.6&15.708&0.010&15.290&0.004&0.043 \\
11&NGC 2543&26.3&15.986&0.011&15.516&0.005&0.069 \\
12&NGC 2537&15.0&14.964&0.007&14.752&0.003&0.054 \\
13&IC 2233&13.7&15.000&0.007&14.805&0.004&0.052 \\
14&IC 2239&88.5&19.177&0.046&18.041&0.016&0.053 \\
\hline
\end{tabular}
\end{center}
\raggedright
{\it Notes:} 1. distances from Paper~I based on $H_0 = 73$\,\kmsmpc \\
 2. AB magnitude \\
 3. Milky Way colour excess in magnitudes from \citet{schlegel98}.
\end{table*}

\subsection{Infrared, Radio, and Visible data}
\label{oir}

The Two Micron All Sky Survey (2MASS), {\it Infrared Astronomical
Satellite} {\IRAS}, and {\it Spitzer}/Infrared Array Camera (IRAC)
imaging data are described in Paper~I. In this paper we follow
\citet{helou88} and calculate the FIR flux of
galaxies as:
\begin{equation}
F_{\rm FIR}=1.26\times10^{-14}(2.58f_{60}+f_{100})\quad
\rm(W m^{-2})
\label{lfir}
\end{equation}
where $f_{60}$ and $f_{100}$ are the \IRAS\ 60 and 100\m\ flux
densities in units of Jy.  Equation~\ref{lfir} is based directly on
observed flux densities with no extrapolation and represents flux emerging
between 42 and 122\,\micron\ \citep{helou88}.
\new{
The two shortest-wavelength \WISE\ bands are close to \Sp/IRAC bands,
and \WISE\ band~3 is close to \IRAS\ 12\,\micron, so we expect our
results to be directly applicable to SFR measurements from
\WISE. \WISE\ band~4 is close to \Sp/MIPS 24\,\micron, both
interesting for SFR measurements but complicated by the presence of
AGN. The SFRS 
will be useful for future investigation of this wavelength.}

The 1.4\,GHz radio data are all from Paper~I. For most galaxies the
data come from the NVSS \citep{Condon1998} or from deeper
observations taken with the same VLA configuration~(D).

The visible spectroscopic data used in this paper were taken from the
Data Release~13 (DR13) of the Sloan Digital Sky Survey
\citep[SDSS;][]{albareti16} or from the central pixels of long-slit
spectra \citep{Marag2017b}. The SDSS spectra were obtained using two
fiber-fed double spectrographs covering a wavelength range of
3800--9200\,\AA\ with spectral resolving power varying between $1850
< \lambda/\Delta \lambda < 2200$. The 3\arcsec\ fiber spectra are
available for 189 SFRS galaxies, of which 187 have detectable
H$\alpha$ line emission. The long-slit data refer to regions 3\farcs5
by 3\arcsec\ in size and spectral resolving power $\sim$1000.  Full
details are given by \citet{Marag2017b}.

\subsection{Photometry and aperture corrections}

The UV and NIR photometry were measured consistently using
\textsc{SExtractor} \citep{bertin96}. The two {\it{GALEX}} images and
the four IRAC mosaics were first registered using \textsc{swarp}
\citep{bertin02} to bring all six images to a common WCS and pixel
size of 0\farcs867. \textsc{SExtractor} was then run in dual-image
mode with objects detected on the 3.6\m\ {\it{IRAC}} image (see
\S\ref{oir}). The 3.6\m\ image was chosen for the initial reference
because this band is most sensitive to galaxy starlight. All
images were inspected to check whether tidal features visible in UV
and/or IR were included in estimating the total magnitudes.  If not,
the relevant processes above were repeated with a more suitable
detection image. For the excessively UV-bright (XUV) galaxies
\citep{gildepaz07,thilker07} such as NGC~4395, the \nuv image was
used for aperture selection and for total flux estimates via
\textsc{SExtractor} `MAG\_AUTO'\footnote{Following
  \url{http://galex.stsci.edu/gr6/?page=faq} we assumed the effective
  wavelengths ($\lambda_{\rm eff}$) for {\it GALEX} to be 1516\,\AA\ and
  2267\,\AA\ in the FUV and NUV, respectively.  The UV counts per
  second measured by \textsc{SExtractor} `MAG\_AUTO' were converted
  to flux densities at effective wavelengths ($f_\lambda$) using the
  unit response also given on the above webpage. $f_\lambda$ were
  then converted to flux densities $f_\nu$ at the corresponding
  frequencies, and those were translated to AB magnitudes and their
  uncertainties. {The resulting AB magnitudes for one count
  per second are 18.824 and 20.036 in FUV and NUV respectively.}}.

Next we applied the extended-source correction to the UV
fluxes. Figure~4 of \citet{morrissey07} shows that the aperture
correction for apertures of radius $<$3\farcs8 and $>$6\arcsec\ are
approximately linear. Hence, following Figure~4 of
\citet{morrissey07} we binned our data into three sets with
$r<3\farcs8$, $3\farcs8<r<6\arcsec$, and $r>6\arcsec$, where $r$ is
the half-light radius from the 2MASS catalog (\S\ref{oir}). To
the first set we applied a linear correction of the form $Ar+B$,
where $A=-0.5608 (-0.6521)$ and $B=2.4912(3.0682)$ respectively in
FUV(NUV). To the second set we applied the corrections suggested by
\citet{morrissey07} for the 6\arcsec\ radius aperture. To the third
set we applied the linear correction by approximating the curve of
growth from 6\arcsec\ onwards such that $A=-0.0016(-0.0028)$ and
$B=0.14(0.2128)$ respectively for FUV(NUV). 

A recent catalog of GALEX measurements for 4138 nearby galaxies
\citep{Bai2015} includes FUV measurements for 42 and NUV measurements
for 73 SFRS galaxies.  For the galaxies in common, the FUV magnitudes
presented here are in the median 0.03\,mag brighter than the
Bai \etal\ D25 magnitudes and 0.07\,mag fainter than their asymptotic
magnitudes. Corresponding values for NUV are 0.02\,mag brighter and
0.10\,mag fainter respectively. The agreement in the medians shows
that our calibration is on the same scale as that of Bai \etal\
Individual galaxies, however, show differences between our
magnitudes and the Bai \etal\ D25 magnitudes with standard deviations
of 0.29 and 0.23\,mag in the FUV and NUV bands, respectively. These
dispersions represent the effect of different choices of aperture and
perhaps also subtraction of sky background and contaminating
sources. We also compared the UV fluxes obtained for the SFRS
galaxies with those of \citet{dale07} for six galaxies in common with
the SINGS sample \citep{kennicutt03} and found them to agree within
the uncertainties.

\subsection{Milky Way Extinction}
\label{extinction}

Dust in the Milky Way attenuates light from external galaxies. The
degree of extinction depends on {position} and {may} require a
large correction in the UV wavebands. We applied a correction of
$A_{\rm FUV}=8.29\times E(B-V)$ and $A_{\rm NUV}=8.18\times E(B-V)$
\citep{seibert05}, where the {adopted}
colour excess {values} $E(B-V)$ come from the
dust reddening maps of \citet{schlegel98}\footnote{obtained from the
  NASA Extra-galactic Database (NED);
  http://nedwww.ipac.caltech.edu/}. These reddening maps are based on
the reprocessed composite of the {\it COBE}/DIRBE and \IRAS/ISSA maps
at 100\m\ with the zodiacal foreground and confirmed point sources
removed. For the SFRS galaxies, colour excesses are in the range
$0.007 \leq E(B-V) \leq 0.164$ with a median of $0.023$\,mag and are
listed in Table~\ref{t:uv}. For
the subsample of SFRS galaxies analysed here, {95 per cent of the FUV
  corrections are less than 0.45\,mag.
Recent work \citep{Lenz2017} confirms that the uncertainties in
  the corrections are generally negligible for current purposes.}

\section{Star Formation Rate indicators}
\label{s:sfr}

\subsection{Individual Star Formation Indicators}

A measure of SFR wholly independent of emission at any other wavelength
comes from the 1.4\,GHz radio emission. 
This radio emission \citep{condon92} comes mainly from non-thermal 
synchrotron radiation from the relativistic electrons in
the remnants of \new{core collapse} supernovae. There is also  a small
contribution from thermal bremsstrahlung from \ion{H}{ii}
regions \citep{condon92,schmitt06,Tabatabaei2017} and potentially
from an active galactic 
nucleus.  Unfortunately the theoretical ratio of radio
emission to SFR depends on the uncertain mass cutoff for Type~II
supernovae \citep{sullivan01}. Therefore the calibration usually is
taken from an
empirical comparison in the local universe \citep[e.g.,][Eq.~13]{yun01}
of the SFR density $\rho_{\rm SFR}$ to radio power density $U_{\rm
  SF}$ related to star formation (i.e., subtracting radio emission
from active galactic nuclei):
\begin{equation} ({\rm SFR}_{\rm 1.4\,GHz}/\Msol\,{\rm yr^{-1}}) = (5.9
  \pm 1.8) (L_{\rm 1.4\,GHz}/ {10^{22} \rm\, W\,Hz^{-1}})\quad.
\label{radio-sfr}
\end{equation}
This relation was derived by combining an integrated 1.4\,GHz
luminosity density $U_{\rm SF} = 2.4 \times
10^{19}$\,W\,Hz$^{-1}$\,Mpc$^{-3}$ with $\rho_{\rm
  SFR}=0.014\pm0.005$\,\Msol\,yr$^{-1}$Mpc$^{-3}$ (both corrected to
$H_0=73$~km~s$^{-1}$~Mpc$^{-1}$).  These values were derived for the
{\it IRAS} 2\,Jy sample 
of galaxies with $L_{\rm 1.4\,GHz} < 10^{24}$\,W\,Hz$^{-1}$. More
recent values are $U_{\rm SF} = (2.17\pm0.10) \times 10^{19}$
\citep{mauch07} and $\rho_{\rm SFR} = 0.012\pm0.001$ \citep{gd16},
which give almost the same ratio. Other estimates of the $\rho_{\rm
  SFR}$ \new{differ by factors of 0.87--2.25}
\citep{gilbank10}\footnote{\citeauthor{gilbank10}
  and \citeauthor{Tabatabaei2017} based their SFRs on a Kroupa IMF.
  We have divided by 0.67 \citep{Madau2014} to convert to the Salpeter IMF
  used here.} depending on method.  \new{\citet{Tabatabaei2017}
  compared radio SFR to a range of other indicators and found
  calibrations of 0.94--2.0 times that of Equation~\ref{radio-sfr}.}
Purely theoretical
calculations give SFRs 1.4 times \citep{schmitt06} or 2 times
\citep{Tabatabaei2017} higher \new{than  Equation~\ref{radio-sfr}}.
\new{The differences in calibration methods} are the main
uncertainty in Equation~\ref{radio-sfr}.
We adopted \new{the value shown} because of its wide use and
scaling consistent with  other indicators.

\begin{figure*}
\begin{center}
\includegraphics[width=\textwidth,clip=true,bb=54 300 646 672]{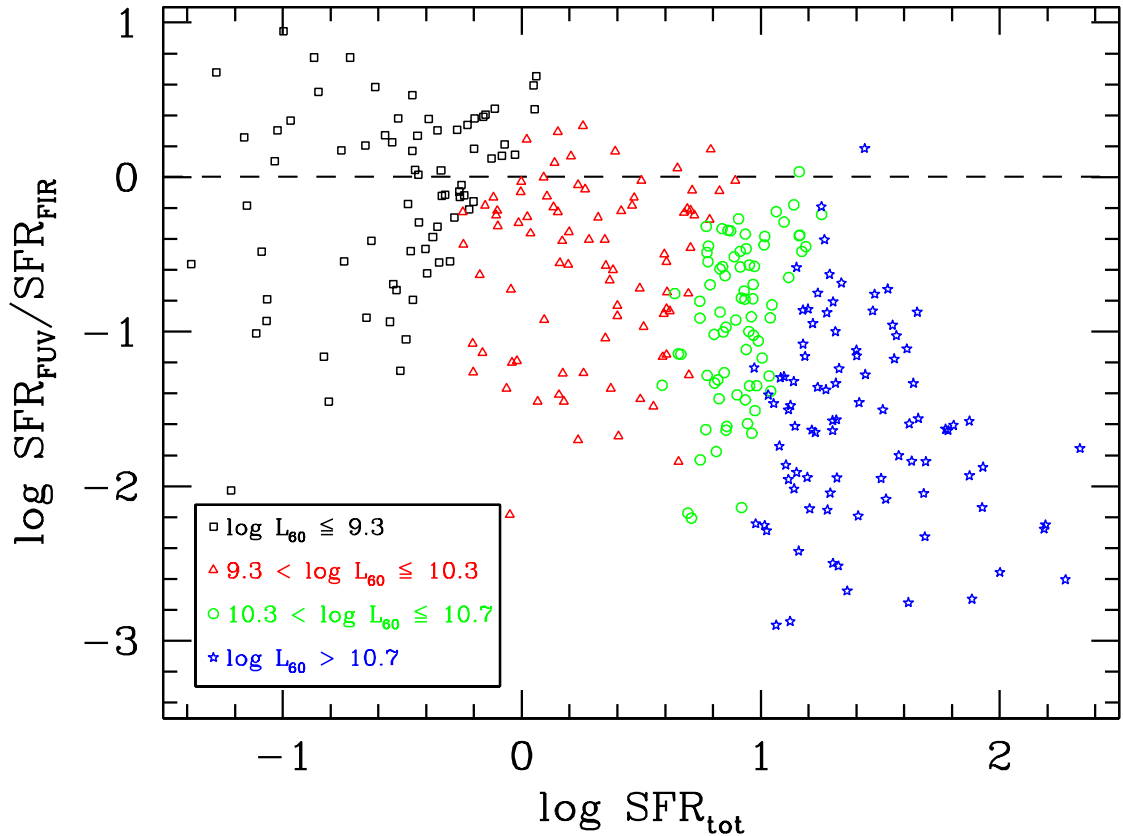}
\end{center}
\caption{Ratio of SFRs calculated solely from FUV photometry to those
  derived solely from FIR photometry as a function of their sum.  The
  symbols indicate individual galaxies coded by quartiles of
  $L_{60\,\micron}$: black squares for $L_{60\,\micron} <
  10^9$~\Lsol, red triangles for $10^9 < L_{60\,\micron} <
  10^{10}$~\Lsol, green circles for $10^{10} < L_{60\,\micron} <
  10^{10.5}$~\Lsol, and blue stars for $L_{60\,\micron} >
  10^{10.5}$~\Lsol. The abscissa is \sfrt\ from
  Equation~\ref{tot-sfr}, and the ordinate is the ratio of SFRs given by
  Equations~\ref{fuv-sfr} and~\ref{fir-sfr}. }
\label{fusfr}
\end{figure*}

A problem with the linear \sfrr\ (Eq.~\ref{radio-sfr}) is that it
tends to be too low at low SFR and too high at high SFR.
\citet{Chi1990} and \citet{bell03} among others have suggested that
in low-luminosity (or low-SFR) galaxies, a fraction of the cosmic
rays accelerated by SNRs may escape from the galaxy. This
scenario might explain the underestimated \sfrr\ of some
low-luminosity galaxies in our sample. \citet{Tabatabaei2017}
suggested that galaxies with high SFR might have stronger magnetic
fields and a flatter energy distribution of relativistic
electrons. This would imply stronger radio emission for a given SFR
at high SFR.  \citet{Davies2017} suggested a non-linear
relation\footnote{Their Eq.~3, here converted to Salpeter IMF by
  multiplying by 1.53.  Their Eq.~2 is similar but has an exponent of
  0.66.  This turns out to give more scatter and a smaller overall
  range than Eq.~3, and therefore we adopt the
  latter. \citealt{Tabatabaei2017} also suggested non-linear
  relations with similar or larger exponents.} based on an
essentially radio-selected sample of nearby galaxies:
\begin{equation}
({\rm SFR2}_{\rm 1.4\,GHz}/\Msol\,{\rm yr^{-1}}) = 5.25
     (L_{\rm 1.4\,GHz}/ {10^{22} \, \rm W\,Hz^{-1}})^{0.75}\quad,
\label{radio-sfr2}
\end{equation}
{and \citet{Brown2017} found a nearly identical relation.}
A non-linear relation as in Equation~\ref{radio-sfr2} could
compensate for cosmic ray escape at low SFR and increased radio
emission at high SFR. Both the linear and non-linear relations are
examined below.

\new{An inevitable output of recent star formation is} UV continuum
emission from 1200--3200\,\AA. \new{UV continuum can be therefore be
  used as an SFR indicator.} We used the prescriptions provided by
\citet{iglesias06}:
\begin{equation}
\log ({\rm \sfrfuv/\Msol\,yr^{-1})}=\log (L_{\rm FUV}/\Lsol)-9.51\quad, 
\label{fuv-sfr}
\end{equation} 
and
\begin{equation}
\log ({\rm \sfrnuv/\Msol\,yr^{-1}})=\log (L_{\rm NUV}/\Lsol)-9.33\quad,
\label{nuv-sfr}
\end{equation}  
where $L_{\rm FUV}$ and $L_{\rm NUV}$ are the intrinsic FUV and NUV
luminosities. \citet{iglesias06} derived their calibrations from
Starburst99 \citep{leitherer99} assuming a solar
metallicity. {\citet{hao11} found SFR 20 per cent larger than given by
  Equation~\ref{fuv-sfr},} and \citet{McQuinn2015} suggested
increasing the \sfrfuv\ calibration by a factor of 1.53.  Neither
will change our results because for the SFRS sample, the
\new{measured UV output} represents only a small part of the total
SFR measurement.  \new{The SFR could be overestimated if an old
  stellar population contributes significant UV emission. Indeed some
  elliptical galaxies, which have low SFR, show a `UV upturn'
  \citep{OConnell1999}. In most galaxies, this is due to residual
  star formation \citep{Yi2011}, but in a small fraction the UV
  emission can come from horizontal branch stars
  \citep[e.g.,][]{Kjaergaard1987}. While such emission will add
  slightly to the observed scatter in SFR relations, the small size
  of the effect even in elliptical galaxies shows that it will be
  negligible for most of the SFRS galaxies because the SFRS selection
  requires dust emission.}

\begin{figure*}
\begin{center}
\includegraphics[width=\textwidth,clip=true,bb=54 244 628 700]{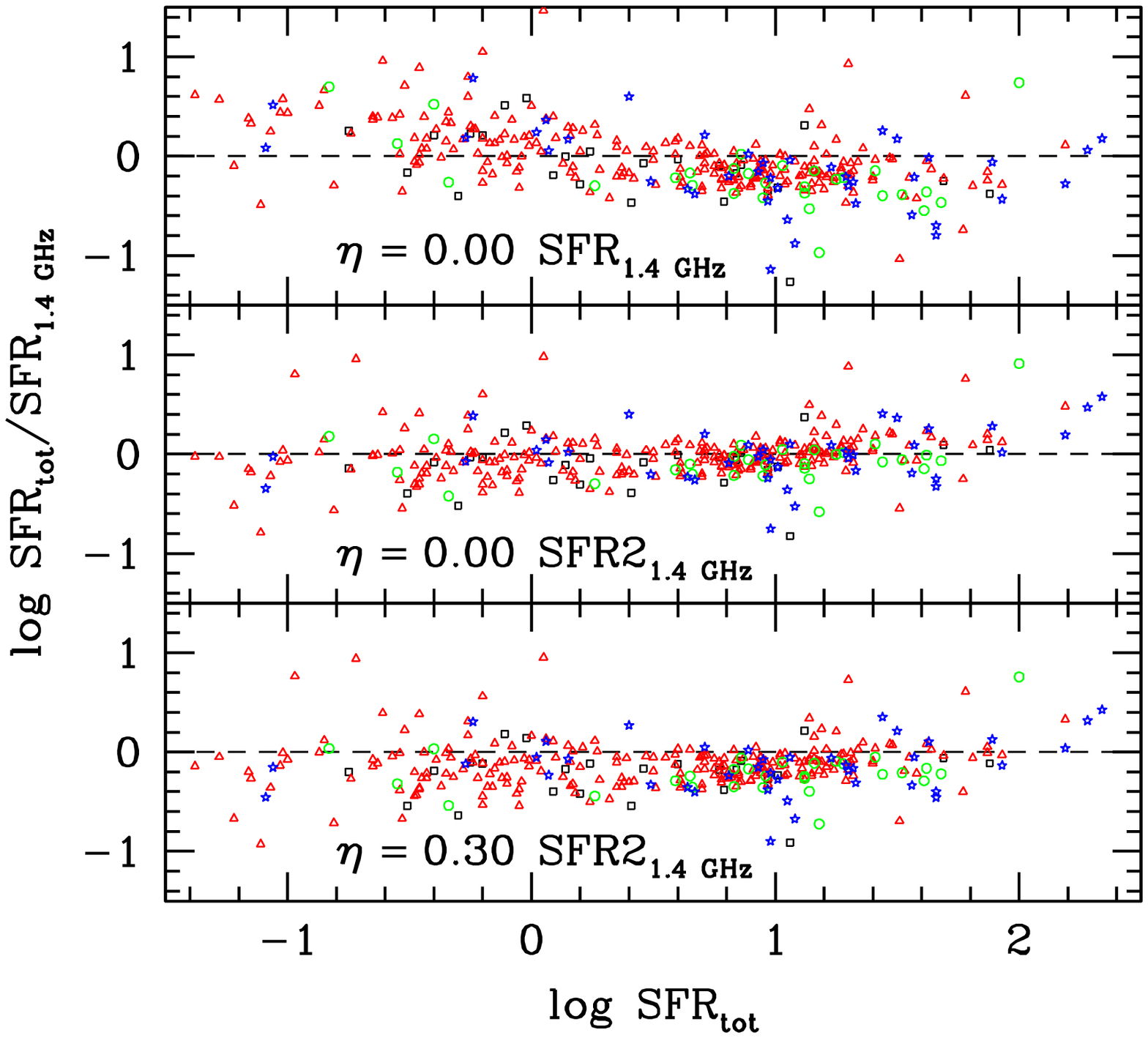}
\end{center}
\caption{Ratio of \sfrt\ (Eq.~\ref{tot-sfr}) to radio SFR as a
  function of \sfrt. {\it(top)} Linear \sfrr\ (Eq.~\ref{radio-sfr})
  and $\eta=0$. {\it(middle)} Non-linear \sfrrd\
  (Eq.~\ref{radio-sfr2}) and $\eta=0.0$. {\it(bottom)} Non-linear
  \sfrrd\ and $\eta=0.3$. The abscissa for all panels uses
  $\eta=0$. Black squares, red triangles, green circles, and blue
  stars represent the four quartiles of 60~\micron\ luminosity as
  indicated in the Figure~\ref{fusfr} legend.  Dashed lines show
  equality of the SFR measures being compared.}
\label{eta}
\end{figure*}

For all but the least dusty galaxies, most of the UV radiation
emitted by young stars is absorbed by dust and reemitted in the FIR.
Therefore, in order to use the measures in Equations~\ref{fuv-sfr}
or~\ref{nuv-sfr}, the observed UV emission has to
be corrected for extinction by one of the methods discussed in
Section~\ref{s:extinc}.  Alternatively, the reemitted FIR radiation itself
can be used as the SFR tracer \citep{kennicutt98}.  Here we adopt the
estimate from \citet{iglesias06}:
\begin{equation}
\log {\rm (\sfrfir}/{\rm \Msol\,yr^{-1}}) =\log (L_{\rm
  FIR}/\Lsol)-9.75\quad, 
\label{fir-sfr}
\end{equation}
where  \lfir\ is based on Equation~\ref{lfir} and the galaxy
distance.

For the SFRS sample, the observed \sfrfuv\ is usually
but not always negligible compared to \sfrfir.
This is evident from Figure~\ref{fusfr}, which also shows
that the \sfrfuv/\sfrfir\ ratio is a strong function of \sfrt.  Of our
326 galaxies with UV photometry, only 47 have $\sfrfuv > \sfrfir$.
This is not surprising because the SFRS sample is based on {\it IRAS}
FIR detections.  Despite that, UV-bright galaxies are represented in
the SFRS sample because of its selection criteria.

\lfir\ derived from Equation~\ref{lfir} differs from `Total Infrared
Luminosity' (\ltir), which takes into account all flux from 3 to
1100\,\micron. \ltir\ has sometimes been used to derive SFR.  $F_{\rm
  TIR}$ can be extrapolated from $F_{\rm FIR}$ and 
$f_{60}/f_{100}$ \citep[Eq.~3 of][]{Dale2001} \new{or can be directly
  measured if \Her/PACS data are available \citep{Galametz2013}.} For
galaxies in our 
sample, \new{the \IRAS\ extrapolation gives} $1.38\le F_{\rm
  TIR}/F_{\rm FIR}\le 11$ with median 2.26 and 
standard deviation 0.26.  \new{Only 56 SFRS galaxies have any
  \Her/PACS data available.  For a few, the PACS 70 or 100\,\micron\
  flux densities are more than a factor of 2 below the corresponding
  \IRAS\ flux density. These galaxies are all notably extended in the
  IRAC images, and presumably the smaller \Her\ beam is not picking
  up all the flux that \IRAS\ saw.  For the 38 SFRS galaxies having
  160~\micron\ PACS data and with PACS 70 or 100~\micron\ data in
  agreement with IRAS, the median $F_{\rm TIR}$ is 0.10~dex higher
  than the IRAS extrapolation, and the dispersion of the ratio is
  0.08~dex. This limited evidence suggests that extrapolating \IRAS\
  flux densities to derive \ltir\ is reasonable}, but \ltir\ is not
used here except where needed to compare with previous results.

The \sfrfir\ measure  suffers from two complications: some
of the UV emission escapes from the galaxy without heating 
dust, and
the FIR emission includes a contribution from dust heated by older
stars {\citep{Sauvage1992}}.  Therefore an improved estimate is
often assumed to be  
\citep{hirashita03}
\begin{equation}
 \rm \sfrt = SFR_{\rm   FUV}+(1-\eta) \sfrfir
\label{tot-sfr}
\end{equation}
where \sfrfuv\ is calculated from the {\it observed} FUV emission,
uncorrected for dust extinction, using
Equation~\ref{fuv-sfr}. {The first} term accounts for UV
radiation that escapes without heating dust, and $\eta$ {in the
  second term} is the fraction of FIR luminosity produced by dust
heated by old stellar populations. (\citealt{Kennicutt2012} and
\citealt{Calzetti2013} have reviewed this subject.) 
Using \lfir\ instead of \ltir\
decreases the effect of dust heated by older stars because such dust is
generally cooler than dust within star formation regions
\citep{Helou1986}. Omitting the $\lambda>122$\,\micron\ radiation
leads to smaller values of $\eta$.

The fraction $\eta$ {of dust luminosity coming from old stars}
can be estimated using evolutionary synthesis models. Earlier
studies involving starburst and luminous IR galaxies found $\eta \sim
0.6$ \citep[\eg][]{buat99, meurer99,gordon00}, but empirical values
for normal star-forming galaxies are much lower
\citep{bell03,hirashita03,buat05,kong04,hao11}, \new{even though} most
studies have used TIR to derive   $\eta$.
For a diverse sample, \citet{bell03} estimated
the contribution of old stellar populations to IR luminosity to be
32$\pm$16 per cent for galaxies  with $L_{\rm TIR}< 10^{11}$\Lsol\ and
9$\pm$5 per cent for LIRGs.  \citet{hirashita03} found
$\eta=0.40\pm0.06$ for a sample of spiral and irregular galaxies in
nearby galaxy clusters but $\eta=-0.04\pm0.09$ (i.e., $\eta\approx0$)
for a sample of starburst galaxies selected at $\sim$1900\,\AA.  For
{\it AKARI}/FIS galaxies observed by SDSS/DR7 and {\it GALEX},
\citet{buat11} found $\eta = 0.17 \pm 0.10$. More recently,
\citet{Boquien2016} studied galaxy regions $\sim$1\,kpc in size and
found $\eta = 0.42$ for the luminosity-weighted average but ranging
from $\sim$0.15 to $\sim$0.7 for different regions.\footnote{Boquien
  \etal\ used different notation than we do, but $k_{\rm TIR}$ in
  their notation corresponds to $(1-\eta)$ in ours.}
Our results presented below are roughly equivalent to 
basing SFR on \ltir\ and using  $\eta=0.56$, the  median value 
  of \ltir/\lfir.  However, using \ltir\ would require either
additional data beyond 100~\micron\ or an uncertain
extrapolation. As will be seen below, our choice to use \lfir\ rather
than \ltir\ is justified by
its success.

\begin{table*}
\begin{center}
\caption{SFR measures\protect\footnotemark[1] for SFRS galaxies
  (complete table available in appendix).}  
\label{tab-sfr} 
\begin{tabular}{rrrrrrrr}
\hline\hline
SFRS\footnotemark[2] & \sfrfuv & \sfrnuv & \sfrr & \sfrrd &
 \sfrfir & \sfrp & \sfrt\footnotemark[3] \\
\hline
1&$-$0.18&0.06&0.97&0.81&0.57&0.74&0.64 \\
2&0.16&0.36&0.89&0.76&0.65&0.84&0.78 \\
3&$-$0.25&$-$0.19&$-$0.63&$-$0.25&$-$0.69&$-$0.53&$-$0.11 \\
5&0.29&0.61&1.53&1.18&1.14&1.22&1.20 \\
8&0.63&0.85&1.30&1.03&0.92&1.31&1.10 \\
9&0.37&0.49&0.70&0.63&0.45&0.45&0.71 \\
10&0.28&0.44&0.69&0.63&0.48&0.68&0.69 \\
11&$-$0.49&$-$0.30&$-$0.22&0.02&$-$0.19&0.03&$-$0.01 \\
12&$-$0.62&$-$0.54&$-$0.78&$-$0.34&$-$0.66&$-$0.77&$-$0.34 \\
13&$-$0.72&$-$0.64&$-$1.58&$-$0.87&$-$1.30&$-$1.77&$-$0.61 \\
14&$-$0.76&$-$0.31&0.84&0.73&0.85&0.66&0.86 \\
\hline
\end{tabular}
\end{center}
\raggedright
{\it Notes:} 1. SFRs in units of $\rm \log(SFR/\Msol\,yr^{-1})$. \\
2. Paper~I \\
3. $\log{\rm (\sfrfir + \sfrfuv)}$
\end{table*}

Comparing the bottom
two panels of Figure~\ref{eta}  shows that 
the main effect of {varying} $\eta$ is to change
the scaling between \sfrr\ (or \sfrrd) and \sfrt\ and that
the {adopted} calibration of
Equation~\ref{radio-sfr2} {gives a preference for}  $\eta\ll0.3$.
Adopting $\eta=0.3$ {in place of $\eta=0$}
would require decreasing the constant in Equation~\ref{radio-sfr2} by
0.12\,dex. This is within the uncertainties of the calibration of
{\sfrr}
\citep[e.g.,][]{yun01,hopkins03,schmitt06,gilbank10,Davies2017}. 
Using any SFR calibration to estimate $\eta$ is of  limited use
because the relative calibrations of different SFR
indicators  can
always be adjusted (within limits) to make the sample median SFRs agree.
As shown in the top two panels of Figure~\ref{eta}, the non-linear prescription
for \sfrr\ {is strongly preferred because it}
removes the tendency for radio luminosity to underestimate
SFR at low SFR values \citep{bell03}.
Leaving aside the overall calibrations,
in principle the {\em scatter} in the
\sfrt/\sfrrd\  ratio can
be used to estimate $\eta$.  
The dispersion of $\sfrrd/\sfrt$ is smallest for $\eta=0$ (0.24\,dex)
but nearly constant with $\eta$ (0.25\,dex for 
$\eta=0.3$). If we restrict the sample to
the central two quartiles of 60\,\micron\ luminosity, the scatter is
0.16\,dex for $\eta=0$ and 0.17\,dex for $\eta=0.3$. 

\begin{table*}
\begin{center}
\caption{Correlation ($y_i=a+bx_i$) comparing different log SFR distributions}
\label{correlations}
\begin{tabular}{crcccccrcccc}
\hline
SFR pairs ($x_i$--$y_i$)& 
\multicolumn{5}{c}{Regression of $y_i$ on $x_i$} & & 
\multicolumn{5}{c}{Regression of $x_i$ on $y_i$} \\ 
\cline{2-6}
\cline{8-12}
 & \multicolumn{1}{c}{\0$a_1$}    & $b_1$      
&   $\sigma_a$\footnotemark[1]  & $\sigma_b$\footnotemark[2] & 
 $\sigma_1$\footnotemark[3]  &  & 
 \multicolumn{1}{c}{\0$a_2$} & $b_2$ & $\sigma_a$\footnotemark[4]  &
 $\sigma_b$\footnotemark[5]  & $\sigma_2$\footnotemark[3]    \\
 \hline  
1.4~GHz$_{\rm NL}$\footnotemark[6]--Total 
 & $-$0.006 & 0.970 & 0.017 & 0.017 & 0.236 &
 & 0.060 & 0.935 & 0.016 & 0.017 & 0.232 \\
PAH--Total 
 & 0.148 & 0.830 & 0.018 & 0.018 & 0.284 &
 & $-$0.086 & 1.042 & 0.022 & 0.023 & 0.318 \\
FIR--Total
 & 0.198 & 0.863 & 0.008 & 0.008 & 0.127 &
 & $-$0.212 & 1.127 & 0.010 & 0.010 & 0.145 \\
PAH--1.4~GHz$_{\rm NL}$
 & 0.182 & 0.811 & 0.019 & 0.019 & 0.289 &
 & $-$0.118 & 1.055 & 0.023 & 0.024 & 0.330 \\
PAH--FIR
 & $-$0.057 & 0.962 & 0.019 & 0.019 & 0.293 &
 & 0.109 & 0.925 & 0.018 & 0.018 & 0.287 \\
1.4~GHz$_{\rm NL}$--FIR
 & $-$0.233 & 1.118 & 0.017 & 0.018 & 0.243 &
 & 0.237 & 0.827 & 0.013 & 0.013 & 0.209 \\
\hline
1.4~GHz--Total
 & 0.132 & 0.727 & 0.015 & 0.013 & 0.236 &
 & -0.109 & 1.247 & 0.021 & 0.022 & 0.309 \\
PAH--1.4~GHz
 & 0.054 & 1.081 & 0.025 & 0.025 & 0.386 &
 & 0.031 & 0.791 & 0.021 & 0.018 & 0.330 \\
1.4~GHz--FIR
 & -0.074 & 0.839 & 0.016 & 0.013 & 0.243 &
 & 0.128 & 1.102 & 0.017 & 0.018 & 0.279 \\
\hline
FUV--Total
 & 0.848 & 0.673 & 0.041 & 0.053 & 0.633 &
 & $-$0.695 & 0.494 & 0.037 & 0.039 & 0.542 \\
NUV--Total
 & 0.711 & 0.841 & 0.033 & 0.052 & 0.576 &
 & $-$0.470 & 0.532 & 0.032 & 0.033 & 0.458 \\
1.4~GHz$_{\rm NL}$--FUV
 & $-$0.691 & 0.467 & 0.039 & 0.041 & 0.560 &
 & 0.848 & 0.614 & 0.042 & 0.054 & 0.642 \\
\hline
1.4~GHz--FUV
 & $-$0.625 & 0.350 & 0.036 & 0.031 & 0.560 &
 & 0.941 & 0.819 & 0.056 & 0.072 & 0.856 \\
1.4~GHz$_{\rm NL}$--NUV
 & $-$0.470 & 0.510 & 0.034 & 0.035 & 0.478 &
 & 0.724 & 0.778 & 0.034 & 0.053 & 0.591 \\
1.4~GHz--NUV
 & $-$0.398 & 0.383 & 0.031 & 0.026 & 0.478 &
 & 0.776 & 1.037 & 0.045 & 0.071 & 0.787 \\
PAH--FUV
 & $-$0.631 & 0.429 & 0.035 & 0.035 & 0.549 &
 & 0.811 & 0.733 & 0.047 & 0.060 & 0.718 \\
PAH--NUV
 & $-$0.403 & 0.464 & 0.030 & 0.030 & 0.466 &
 & 0.661 & 0.920 & 0.038 & 0.059 & 0.657 \\
FUV--NUV
 & 0.209 & 0.909 & 0.008 & 0.010 & 0.125 &
 & $-$0.237 & 1.055 & 0.008 & 0.012 & 0.135 \\
FUV--FIR
 & 0.701 & 0.653 & 0.050 & 0.065 & 0.771 &
 & $-$0.572 & 0.367 & 0.036 & 0.036 & 0.578 \\
NUV--FIR
 & 0.573 & 0.851 & 0.041 & 0.064 & 0.712 &
 & $-$0.345 & 0.413 & 0.031 & 0.031 & 0.496 \\
\hline
\end{tabular}
\end{center}
\raggedright
{$^1$Uncertainty in $a_1$ ($1\sigma$)}\\
{$^2$Uncertainty in $b_1$ ($1\sigma$)}\\
{$^3$Dispersion of sample from best fit relation (dex)}\\
{$^4$Uncertainty in $a_2$ ($1\sigma$)}\\
{$^5$Uncertainty in $b_2$ ($1\sigma$)}\\
{$^6$Non-linear relation given by Eq.~\ref{radio-sfr2}}\\
\end{table*}

The $\eta=0$ value implied by the scatter in \sfrt/\sfrrd\ differs
from previous results. Our use of FIR rather than TIR is {a
  principal but} probably not the only reason for this: using the
colour-dependent values of $\eta$
suggested by \citet[][their Fig.~6 and any of several colours from FUV
to 3.6\,\micron]{Boquien2016} does not decrease the observed scatter.
{In fact,} the {calculated} rms scatter in $\log{\rm(TIR/FIR)}$ for
our sample is only 0.11\,dex (Sec.~\ref{oir}). The real scatter is
probably higher because the TIR/FIR ratio is based on an
extrapolation using simple dust models \citep{Dale2001}, but the true
TIR/FIR ratio
cannot be evaluated without more observations at wavelengths longer than
100\,\micron.  Better estimates of $\eta$ will require either much
larger samples, better theoretical knowledge of the relative
calibrations of SFR measures, or a new way of estimating values of
$\eta$ for individual galaxies. In the following we use $\eta=0$ and
non-linear \sfrrd\ (Eq.~\ref{radio-sfr2}) because these minimize
{both} the observed scatter and the calibration offsets.

The last SFR measure we consider here is polycyclic aromatic
hydrocarbon (PAH) molecular emission features. (See
\citealt{calzetti10} for a review.) PAHs can form in galaxies from
evolved stars, stellar mass loss, \new{gas cloud collisions}, or
cooling flows and 
are excited by UV emission over a wide range of wavelengths. The PAH
emission arises from photo-dissociation regions, which often surround
\citep{Helou2004} the \hii\ regions that mark the locations of
massive stars. This makes PAH emission an indirect but still useful
SFR tracer.  The 8.0\m\ IRAC band detects a complex of PAH features
in low-redshift galaxies, and \citet{wu05} showed that the 8.0\m\
dust luminosity correlates well with the 1.4\,GHz and the 24\m\
luminosity, both of which are star formation tracers. The 8.0\m\ dust
luminosity also correlates linearly with the MIPS 160\m\ luminosity
\citep{zhu08} and non-linearly with the extinction-corrected
Paschen-$\alpha$ luminosity \citep{calzetti07}. \citet{calzetti07},
\citet{zhu08}, \citet{kennicutt09}, \citet{Shipley2016}, and
\citet{Marag2017a} among others have used PAH emission to estimate
the SFR for galaxies, and \citeauthor{Shipley2016} showed that of the
various PAH features, the one at 7.7\,\micron\ correlates best with
SFR as measured by their combination of 24\,\micron\ and H$\alpha$ emission.

For a majority of local galaxies seen by IRAC, the PAH emission dominates
the 8.0\m\ band \citep{pahre04}. However, a stellar continuum is still present
especially in the early-type galaxies \citep{Helou2004,wu05,huang07}.
To correct for this, we subtracted 0.227 times the 3.6\m\ flux
density from the observed 8.0\m\ flux density to yield the 8.0\m\ flux
density attributable to dust \citep{huang07}.\footnote{This factor is
  close to those used elsewhere
  \citep[\eg][]{Helou2004,wu05,marble10}.} To convert the 8.0\m\ dust
luminosity to SFR, we have used the prescription of \citet{wu05}
\citep[also see][]{zhu08}:
\begin{equation}
{\rm SFR_{\rm PAH}}/\Msol\,{\rm yr^{-1}} = \frac{\nu L_\nu{\rm
    [8.0\,\micron, dust]}}{1.57 \times 10^9\,{\rm L_\odot}}\quad
\label{pah-sfr}
\end{equation}
with $L_\nu{\rm [8.0\,\micron, dust]}$ derived from the IRAC
8\,\micron\ flux density attributable to dust. \citet{Brown2017}
found a non-linear relation between SFR and \lpah\ (analogous to
Equation~\ref{radio-sfr2} for 1.4\,GHz).  That would give a small
overall decrease in dispersion (from 0.32\,dex to 0.28\,dex),
improving the fit mainly for $\sfrt \la 0.5$\,\Msol\,yr$^{-1}$.

\begin{figure*} 
\centering{{\rotatebox{270}{\epsfig{file=sfr-grid-3.ps,width=13.0cm}}}}
\caption{Comparison of SFR measures. Points represent SFRS galaxies
  with point types indicating quartiles of 60~\micron\ luminosity as
  indicated in the legend. Dotted lines show equality of the
  respective SFR measures.  Dot-dash lines show the best-fit
  relations. Histograms show the distribution of each SFR measure as
  indicated under the respective histogram with colours indicating the
  four quartiles of 60~\micron\ luminosity.}
\label{triangle}
\end{figure*}

\begin{figure} 
\begin{center}
\includegraphics[width=0.45\textwidth,clip=true,bb=84 236 584 696]{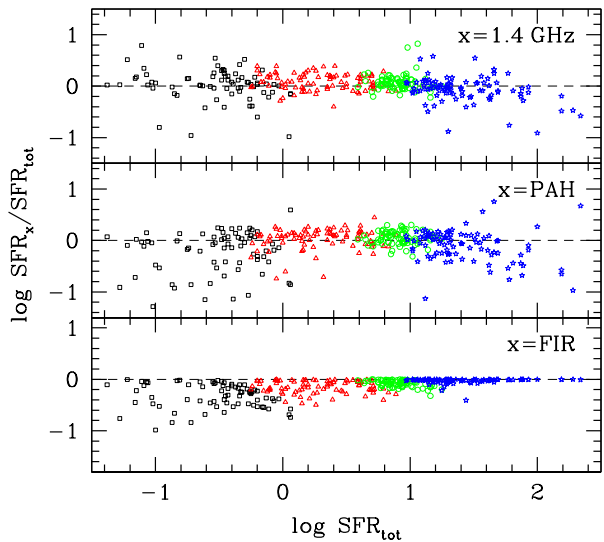}
\end{center}
\caption{The ratio of SFR measured using {{(top to bottom)}}
   1.4\,GHz radio (non-linear measure as given by
   Eq.~\ref{radio-sfr2}), PAH (Eq.~\ref{pah-sfr}), and FIR
   (Eq.~\ref{fir-sfr}) to the total SFR (Eq.~\ref{tot-sfr})
   respectively. Symbols represent the four quartiles of 60~\micron\
   luminosity as indicated in the Figure~\ref{fusfr} legend, and the
   dashed line corresponds to equality.}
\label{frac-sfr}
\end{figure}

\begin{table}
\begin{center}
\caption{Statistical properties of \new{SFR distributions of SFRS
    galaxies} using 
  different  \new{SFR measures} (all in $\log[{\rm SFR}/\Msol\,\rm yr^{-1}]$)}
\label{dist}
\begin{tabular}{crccr}
\hline\hline
 SFR   & Mean & Std. deviation  & Skewness & Median   \\
\hline
  FUV  		& $-$0.42 &  0.67 & $-$0.57 & $-$0.36 \\
  NUV   	& $-$0.17 &  0.62 & $-$0.56 & $-$0.12 \\
  U             & 0.19  &   0.97  &  $-$1.14  &  0.34   \\
  PAH   	&    0.51 &  0.87 & $-$0.75 &    0.77 \\
  FIR    	&    0.43 &  0.88 & $-$0.43 &    0.65 \\
1.4~GHz\footnotemark[1]&    0.60 &  1.01 & $-$0.54 &    0.88 \\
1.4~GHz$_{\rm NL}$\footnotemark[2] 	&    0.59 &  0.76 & $-$0.54 &    0.80 \\
Total   	&    0.57 &  0.77 & $-$0.33 &    0.76 \\
\hline
\end{tabular}
\end{center}
\raggedright
{$^1$Linear relation given by Eq.~\ref{radio-sfr}}\\
{$^2$Non-linear relation given by Eq.~\ref{radio-sfr2}}
\end{table}

\begin{table*}
\begin{center}
\caption{Results for statistical tests between different SFR distributions}
\begin{tabular}{ccccrcccccccc}
\hline\hline
SFR &  \multicolumn{2}{c}{$F$\footnotemark[1]} & &
\multicolumn{2}{c}{Student's $t$\footnotemark[2]} && 
\multicolumn{2}{c}{K-S} && \multicolumn{2}{c}{Pearson's $r$} \\
\cline{2-3}
\cline{5-6}
\cline{8-9}
\cline{11-12}
Distributions & $F$   & prob &&
 \multicolumn{1}{c}{$t$}  & prob    && $D$  & prob     && $r$  & prob \\
\hline
1.4~GHz$_{\rm NL}$\footnotemark[3]--Total 	&  1.337  &  0.009  &  &  $-$0.023  &  0.982  & &  0.089  &  0.144  &  &  0.952  &  0.000   \\
PAH--Total   		&  1.261  &  0.037  &  &  $-$0.922  &  0.357  & &  0.056  &  0.679  &  &  0.933  &  0.000   \\
FIR--Total  		&  1.307  &  0.016  &  &  $-$2.144  &  0.032  & &  0.107  &  0.043  &  &  0.986  &  0.000   \\
PAH--1.4~GHz$_{\rm NL}$ 	&  1.687  &  0.000  &  &  $-$0.956  &  0.339  & &  0.100  &  0.074  &  &  0.927  &  0.000   \\
PAH--FIR     		&  1.036  &  0.751  &  &     1.164  &  0.245  & &  0.079  &  0.247  &  &  0.945  &  0.000   \\
1.4~GHz$_{\rm NL}$--FIR   	&  1.748  &  0.000  &  &     2.250  &  0.025  & &  0.156  &  0.001  &  &  0.961  &  0.000   \\
\hline
1.4~GHz\footnotemark[4]--Total 		&  1.717  &  0.000  &  &     0.485  &  0.628  & &  0.126  &  0.010  &  &  0.952  &  0.000   \\
PAH--1.4~GHz 		&  1.361  &  0.006  &  &  $-$1.266  &  0.206  & &  0.127  &  0.009  &  &  0.927  &  0.000   \\
1.4~GHz--FIR   		&  1.314  &  0.014  &  &     2.332  &  0.020  & &  0.156  &  0.001  &  &  0.961  &  0.000   \\
\hline
FUV--Total  		&  1.326  &  0.011  &  & $-$17.460  &  0.000  & &  0.540  &  0.000  &  &  0.565  &  0.000   \\
NUV--Total  		&  1.546  &  0.000  &  & $-$13.400  &  0.000  & &  0.479  &  0.000  &  &  0.657  &  0.000   \\
1.4~GHz$_{\rm NL}$--FUV   	&  1.008  &  0.939  &  &    18.844  &  0.000  & &  0.546  &  0.000  &  &  0.526  &  0.000   \\
1.4~GHz--FUV   		&  2.276  &  0.000  &  &    15.220  &  0.000  & &  0.546  &  0.000  &  &  0.526  &  0.000   \\
1.4~GHz--NUV   		&  2.654  &  0.000  &  &    11.706  &  0.000  & &  0.506  &  0.000  &  &  0.621  &  0.000   \\
1.4~GHz$_{\rm NL}$--NUV   	&  1.156  &  0.191  &  &    14.536  &  0.000  & &  0.506  &  0.000  &  &  0.621  &  0.000   \\
PAH--FUV     		&  1.673  &  0.000  &  &    15.280  &  0.000  & &  0.553  &  0.000  &  &  0.573  &  0.000   \\
PAH--NUV     		&  1.951  &  0.000  &  &    11.420  &  0.000  & &  0.485  &  0.000  &  &  0.667  &  0.000   \\
FUV--NUV     		&  1.166  &  0.167  &  &  $-$5.008  &  0.000  & &  0.181  &  0.000  &  &  0.980  &  0.000   \\
FUV--FIR     		&  1.733  &  0.000  &  & $-$13.840  &  0.000  & &  0.503  &  0.000  &  &  0.478  &  0.000   \\
NUV--FIR     		&  2.021  &  0.000  &  &  $-$9.974  &  0.000  & &  0.433  &  0.000  &  &  0.581  &  0.000   \\ 
\hline
\label{tabstats}
\end{tabular}
\end{center}
\raggedright
{\it Notes--}`prob' is the probability of the data under the null
hypothesis.\\
{$^1$The statistic $F\!=\!\sigma_1^2/\sigma_2^2$ where $\sigma_i^2$
  is the variance.}\\
{$^2$The statistic
  $t\!=\!(\overline{x_1}\!-\!\overline{x_2})/
  \sqrt{(\sigma_1^2/n_1)\!+\!(\sigma_2^2/n_2)}$
  where $\overline{x_i}$ is the mean of a distribution of $n_i$ data
  points and variance $\sigma_i^2$.} \\
{$^3$Non-linear relation given by Eq.~\ref{radio-sfr2}}\\
{$^4$Linear relation given by Eq.~\ref{radio-sfr}}
\end{table*}

All the SFR measures examined here are given in
Table~\ref{tab-sfr}. For the majority of the sample, the statistical
measurement uncertainty (i.e., from photon and detector noise) is
below the systematic errors.  For radio continuum imaging the
calibration uncertainty is 3 per cent while the systematics add a further
uncertainty of $0.45 \sqrt{n}$\,mJy to extended sources, where $n$ is
the number of beams, each 45\arcsec\ in diameter, covering the
source. For the SFRS galaxies, $n\le5.7$ \citep{condon91}. The
{\it{IRAS}} survey includes a uniform calibration for point sources
of better than 10 per cent over nearly the entire sky \citep{soifer87}. For
the {{IRAC}} and {\it{GALEX}} data, the calibration uncertainty is of
the order of 3 per cent or better.  Despite our efforts to choose apertures
so as to give consistent total magnitudes, aperture uncertainties are probably
at least 0.1\,mag.  {As will be seen in Section~\ref{total}, the
uncertainties in the empirical relations used to convert luminosity
to SFR are larger than these observational uncertainties.}

\subsection{Calibration of Individual SFR Tracers}
\label{total}

The global {radio and infrared} SFR measures are in agreement
with each other as 
shown in Figure~\ref{triangle}. This is the case despite the
prescriptions being based on very different underlying physics and
their calibrations having been {established} from different
samples.  The uncorrected UV measures {correlate with the others
  but} are too low, especially at
large SFR, as expected when extinction is ignored. The respective
correlations, the standard deviation in fitted parameters, and the
goodness of fit parameter $\chi^2$ are listed in
Table~\ref{correlations}. Table~\ref{dist} lists some statistical
properties of the different SFR distributions, and
Table~\ref{tabstats} gives results of statistical tests comparing
them.

Figure~\ref{frac-sfr} compares the indicators that are unaffected by
extinction.  The non-linear \sfrrd\ shows an overall slope that would
become steeper if  the exponent in Equation~\ref{radio-sfr2}
were made smaller (such as 0.66 in Eq.~2 of \citealt{Bai2015}). \sfrp\ is in
good agreement with \sfrt\ for most 
galaxies, but about 18 per cent of galaxies are outliers (8 with
$\log(\sfrp)-\log(\sfrt)>0.3$ and 51 with $\log(\sfrp)-\log(\sfrt)<-0.3$).  \sfrfir\ is in
good agreement with \sfrt\ at high SFR, but it is up to $\sim$1\,dex
low at low SFR because this measure neglects UV light that escapes the galaxy
without heating dust.  This light is most important in low-SFR
galaxies (Fig.~\ref{fusfr}).

Table~\ref{dist} confirms that the \sfrrd, \sfrp, \sfrt\
distributions are comparable in mean and median.  This is also
consistent with the $t$-test (Table~\ref{tabstats}), which checks against
the hypothesis that two distributions with different variances have
the same mean. The K-S statistic (Table~\ref{tabstats}) is consistent
with all three being drawn from the same parent distribution. The
mean and median for \sfrfir\ are a little lower than the previous
three because \sfrfir\ neglects escaping UV light.  Nevertheless, \sfrfir\
is consistent with having been drawn from the same distribution of
SFRs.

The ranges of SFR indicated by \sfrfir, \sfrp, and \sfrt\ are also
similar as indicated by the respective sample standard deviations.
The $F$-test measures the probability that two samples drawn
from a single population would have variances differing by as much as
the observed amount.   The statistic $F$
(Table~\ref{tabstats}) is the ratio of the two variances, and hence a
value $\ll$1 or $\gg$1 indicates significantly different
variances. Table~\ref{tabstats} shows that the differences are at
most marginally significant. In contrast, the linear \sfrr\ shows a
wider range than \sfrt, consistent with its underestimating low SFR and
overestimating high SFR.  \sfrrd\ compensates for that (and would
overcompensate if the exponent in Equation~\ref{radio-sfr2} were made
smaller). 

As expected, the distributions of the UV SFRs differ
from all others because the UV has not been corrected for extinction.
Table~\ref{dist} confirms that the mean of the UV SFRs is significantly
below the monochromatic SFRs estimated at longer wavelengths.

\begin{figure} 
\begin{center}
\includegraphics[width=0.48\textwidth,clip=true,bb=78 300 596 690]{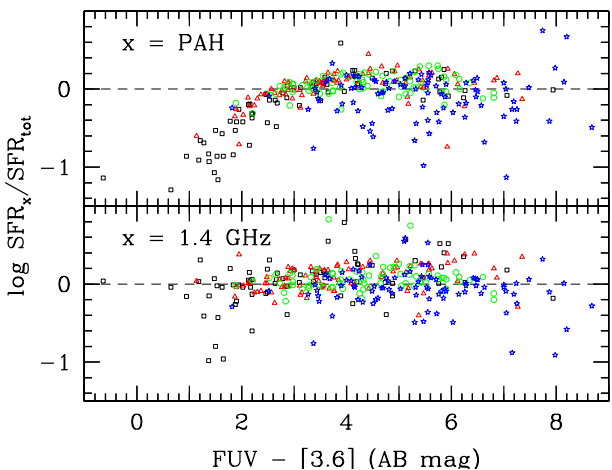}
\end{center}
\caption{Ratio of inferred SFRs as a function of emerging UV
  radiation. The abscissa is the FUV to 3.6\,\micron\ colour as
  measured by IRAC in the same aperture used for the FUV
  magnitude. Upper panel shows \sfrp\ (Eq.~\ref{pah-sfr}, and lower
  panel shows non-linear \sfrrd\ (Eq.~\ref{radio-sfr2}).  Point types
  show quartiles of 60\,\micron\ luminosity as indicated in the
  Figure~\ref{fusfr} legend.}
\label{eta3}
\end{figure}

The good agreement between \sfrp\ and \sfrt\ was also reported by
\citet{Marag2017a}.  As noted above, most large deviations are in the
sense that PAH underestimates the SFR, and most deviant galaxies are
at the extreme ends of the luminosity distribution. The lack of PAH
emission in low-luminosity galaxies has been documented previously
\citep{boselli98,hogg05} and  attributed to lack of PAH grains
\citep[][among others]{wang96,hopkins01,buat05}. The low-luminosity
galaxies can have deficient PAH emission if the galaxies are low in
metallicity, i.e., 
lack the raw material to form PAHs, or if they are too young to have
formed PAH yet. However, given that the shallow potential wells in these
galaxies are unable to retain SNe ejecta for a prolonged duration,
low metallicity might seem to be the most likely cause for the
underestimated \sfrp\ for these galaxies
\citep[\eg][]{leroy06}, and indeed \citet{Shipley2016} confirmed that
galaxies with low metallicity have low PAH emission. However,
Figure~\ref{eta3} shows that PAH 
deficiency is greatest in galaxies exhibiting strong UV radiation
fields, suggesting that PAH destruction may be important.  However,
low metallicity could still be the underlying cause if the reason for the
strong UV radiation field is low dust abundance.  The high-luminosity
galaxies may have a relatively intense radiation field that destroys
the PAHs \citep[\eg][]{condon92}, but there is no evidence for that
in the {\em emerging} UV radiation as seen in
Figure~\ref{eta3}. 
Because of dust extinction, however, the emerging UV radiation may
not be characteristic of the local radiation fields where stars are forming.

The bottom panel of Figure~\ref{frac-sfr} also shows the correlation
between \sfrfir\ and \sfrt. As \sfrt\ increases, the fraction of SFR
traced by \sfrfir\ increases markedly.  For galaxies in the top
luminosity quartile of the SFRS, virtually all the star formation is
traced by FIR emission.  For galaxies in the lowest luminosity
quartile, \sfrfir\ can underestimate \sfrt\ by almost an order of
magnitude for some galaxies, but for other galaxies with the same
SFR, the \sfrfir\ is the dominant contributor.  Figure~\ref{fusfr}
is a more direct demonstration of the importance of escaping UV
radiation as a function of luminosity or SFR.

\section{Extinction Indicators}
\label{s:extinc}

For a galaxy forming stars, the intrinsic UV luminosity is
proportional to the SFR (\citealt{kennicutt98} and references
therein). Dust, depending on its amount and
distribution, absorbs some fraction of the UV and reradiates the
energy in the FIR. If the extinction could be measured, the {\it
  corrected} UV flux would measure the total SFR.

In general, there are two types of extinction indicators for
galaxies. One type is based on the ratio of FIR to UV luminosity
($\equiv$IRX).  Such a `bolometric' extinction indicator in effect
gives a measure of total SFR as in Equation~\ref{tot-sfr} but with a
different {formula to translate from observed
flux densities to SFR}. As for any method involving FIR
emission, the measure is imperfect because older stars can also heat
dust and also because our specific line of sight to a star forming
region may not represent the average over all directions around that
region {because of both galaxy inclination and morphology of the
  dust distribution}. The second type of method uses visible or UV
spectral slope 
($\beta$)
\citep[\eg][]{meurer99,kong04,Cortese2006,gilbank10,overzier11}, Balmer
decrement, or a similar measure of reddening, which is translated to
extinction by means of a chosen reddening curve.\footnote{The UV
  spectral slope $\beta$ is defined by a 
  power-law fit of the form $F_\lambda\propto\lambda^\beta$.} Such
`colour-excess' (or `reddening')
methods are the only choice when FIR data are not available. A major
problem is that colour excess depends critically on the dust geometry
relative to the emitting stars (\ie\ `foreground screen' or `mixed
slab' approximations {or `discrete clouds' or a combination}---see
\citealt{charlot00} for discussion and modeling). Despite the
complications, \citet{meurer99} \citep[also
see][]{Cortese2006,overzier11} found 
an empirical relation between UV colour $\beta$ and IRX that gave rms
scatter 0.3\,dex in IRX in their UV-selected sample of local
galaxies.  {The SFRS allows us to test how well such colour-excess
  methods work in a more representative sample.}

\begin{figure}
\begin{center}
\includegraphics[width=0.48\textwidth,clip=true,bb=66 372 622 676]{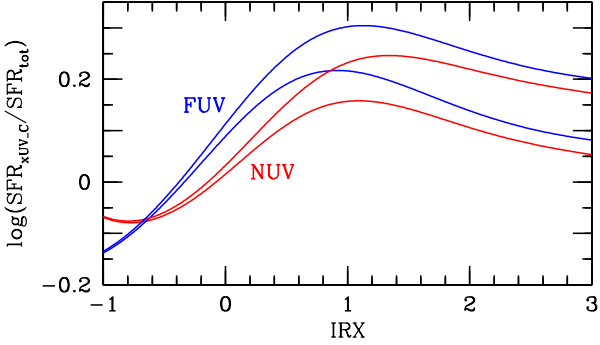}
\end{center}
\caption{Ratio of SFR as calculated by empirical prescriptions
  \citep{buat05} to \sfrt.  {The abscissa IRX is the observed
    total infrared $F_{\rm TIR}$ to UV $\nu F_\nu$ flux ratio, making
    the ordinate depend on $F(60)/F(100)$.} Blue curves show the
  ratio for FUV (Eq.~\ref{afuv}) and 
  the red curves for NUV (text footnote~\ref{fn:anuv}) {for the
    16th and 84th percentiles of $F(60)/F(100)$ of sample galaxies.}}
\label{fig:buat}
\end{figure}

\begin{figure}
\begin{center}
\includegraphics[width=0.48\textwidth,clip=true,bb=90 276 524 718]{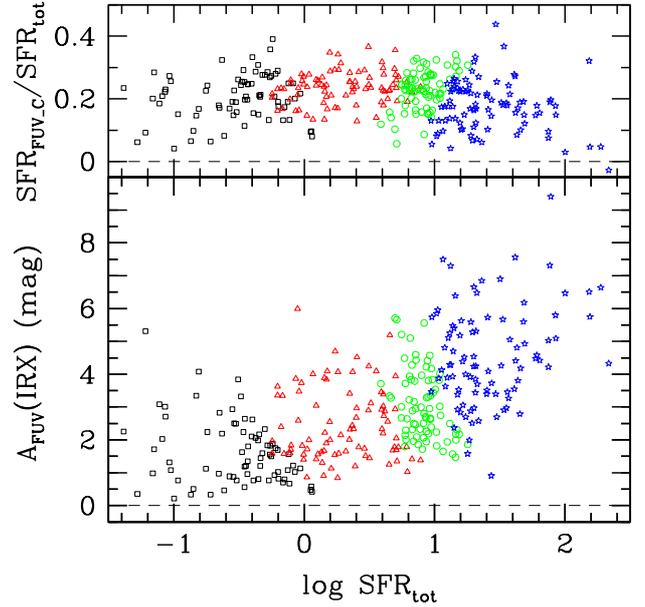}
\end{center}
\caption{FUV extinction and resulting SFR as a function of
  \sfrt. {\it (Bottom:)} FUV bolometric extinction $A_{\rm FUV}(IRX)$
  \citep[Eq.~\ref{afuv}, original source][]{buat05}. Dashed line
  shows zero extinction. {\it (Top:)} Ratio of SFR computed from FUV
  corrected by bolometric extinction {(Eq.~\ref{afuv})} to
  \sfrt. Dashed line shows equality. }
\label{fig:afuv}
\end{figure}

An empirical prescription \citep{buat05} for the bolometric
extinction $A_{\rm  FUV}({\rm IRX})$ derived from IRX is
\begin{equation}
A_{\rm  FUV}({\rm IRX})=-0.0333p^3+0.3522p^2+1.1960p+0.4967~~,
\label{afuv}
\end{equation}
where $p\equiv\log(L_{\rm TIR}/L_{\rm
  FUV})$.\footnote{\label{fn:anuv} The corresponding equation for NUV
  is $A_{\rm NUV}=-0.0495q^3+0.4718q^2+0.8998q+0.2269$ where
  $q\equiv\log(L_{\rm TIR}/L_{\rm NUV})$. Results for FUV and NUV are
  similar, so we discuss in detail only the former. { IRX in
    these equations is based on $F_{\rm TIR}$, not $F_{\rm
      FIR}$. }}
Figure~\ref{fig:buat} shows how this empirical prescription
translates to \sfrt\ for different values of IRX, and
Figure~\ref{fig:afuv} (upper panel) shows that \sfrfuv\ corrected for
extinction by a {\em bolometrically derived} factor is close to $\rm
SFR_{tot}$ for the SFRS sample, {showing the ${\approx}0.2$\,dex
  over-correction expected from Figure~\ref{fig:buat}.} The more
luminous galaxies, those with $L_{60}>10^{10.5}$\,\Lsol, tend to have
larger IRX (as expected from Fig.~\ref{fusfr}) and therefore larger
bolometric extinction. Figure~\ref{fig:afuv} (lower panel)
illustrates this relationship.

\begin{figure}
\begin{center}
\includegraphics[width=0.48\textwidth,clip=true,bb=72 224 536 700]{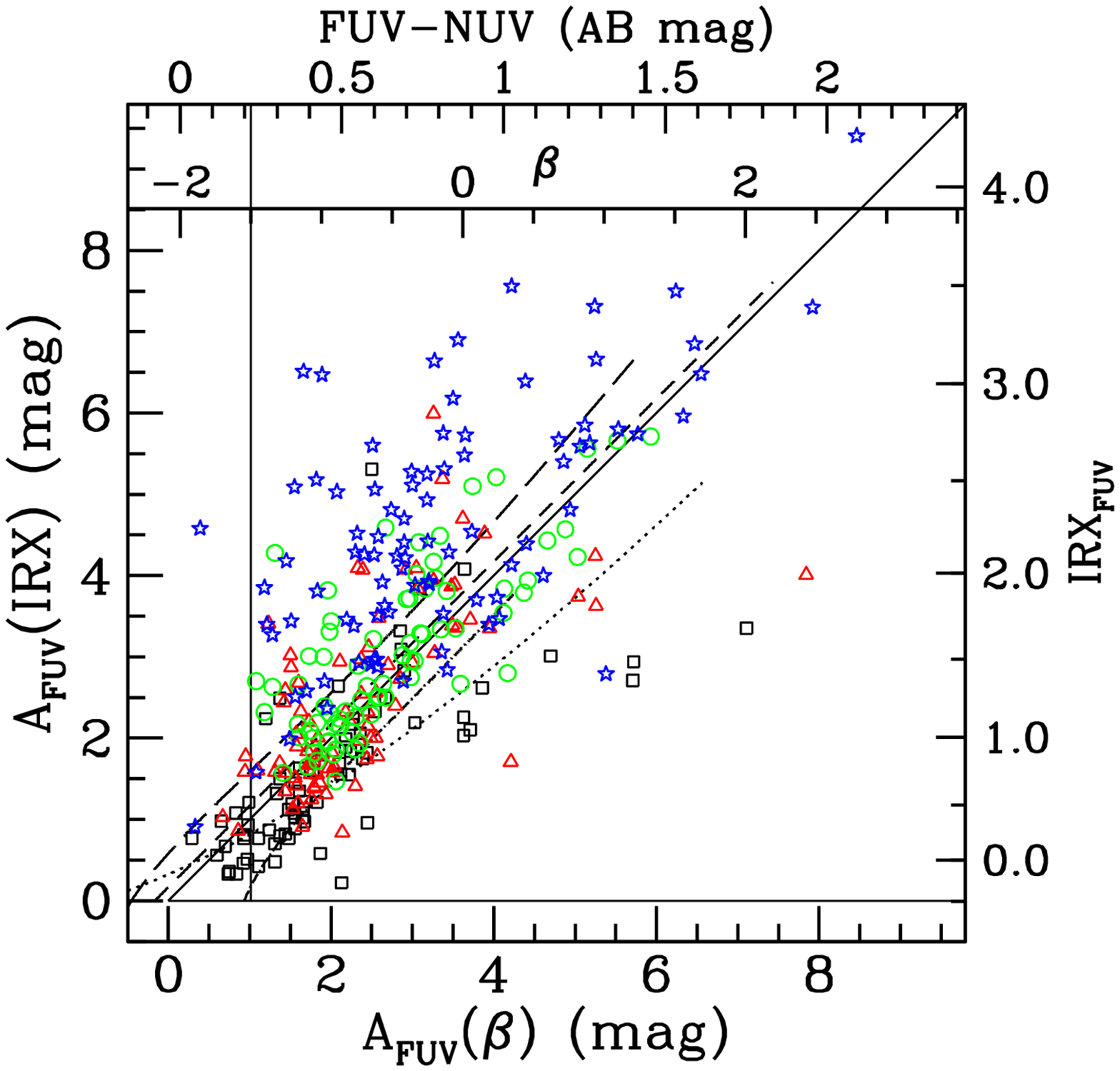}
\end{center}
\caption{Bolometric versus colour-excess FUV extinctions for SFRS
  galaxies.  Measured values of $\rm 
  FUV-NUV$ {\it GALEX} colours and $\beta$ are shown on the upper
  axes. The abscissa 
  is the resulting colour-excess extinction derived from the \citet{seibert05}
  relation.   The ordinate is $A_{\rm FUV}(IRX)$ \citep[Eq.~\ref{afuv},
  original source][]{buat05}; corresponding values of IRX$_{\rm FUV}$
  are shown on the right. The solid line shows equality between
  the two $A_{\rm FUV}$ measures.  The short-dashed diagonal line
  shows the empirical 
  relation from \citet{seibert05}, the long-dashed line shows the
  \citet{kong04} relation (which is almost the same as the
  \citealt{meurer99} relation), the dotted line shows the
    \citet{Cortese2006} relation, and the dash-dot line shows the
    \citet{Casey2014} relation. Lines extend over the range of the
    respective samples.   Point types
  show quartiles of 60\,\micron\ luminosity as indicated in the
  Figure~\ref{fusfr} legend.  {The horizontal solid line
    indicates zero bolometric extinction, which occurs at $\log(\rm IRX)\approx
    -0.49$. The vertical solid line indicates $\beta=-1.5$, which
    corresponds to $A_{\rm FUV}(\beta)=1.05$.}}
\label{fnuv-irx}
\end{figure}

An empirical expression for FUV extinction based on UV reddening of a
diverse, UV-selected sample of 200 galaxies \citep{seibert05} is
\begin{equation}
A_{\rm FUV}(\beta)=3.978(m_{\rm FUV}-m_{\rm NUV})+0.143~~,
\label{eq:beta}
\end{equation}
where $m_{\rm FUV}$ and $m_{\rm NUV}$ are the respective {\it GALEX}
AB magnitudes.\footnote{For the adopted {\it GALEX} effective
  wavelengths, $\beta = 2.289(m_{\rm FUV}-m_{\rm NUV})-2$.} The
equation is similar to relations derived by others
\citep[e.g.,][]{hao11}. Figure~\ref{fnuv-irx} shows the SFRS galaxies
in the IRX--$(m_{\rm FUV}-m_{\rm NUV})$ (or equivalently
IRX--$\beta$) space. There is a correlation between $A_{\rm
  FUV}(\beta)$ and $A_{\rm FUV}({\rm IRX})$ with Pearson correlation
coefficient $r=0.71$ and mean $\langle A_{\rm FUV}(\beta)- A_{\rm
  FUV}({\rm IRX})\rangle = 0.33$\,mag, but the rms scatter in $A_{\rm
  FUV}({\rm IRX})$ as derived from $A_{\rm FUV}(\beta)$ is 0.44\,dex.
Galaxies with $A_{\rm FUV}(\beta)\la2$ can have bolometric extinctions as high
as 6\,mag, and $A_{\rm FUV}(\beta)$ applied to $L_{\rm FUV}$ greatly
underestimates their FIR luminosity and therefore SFR.  This is
consistent with other results
\citep[e.g.,][]{kong04,johnson06,johnson07}, which have shown that
galaxies having higher current SFR relative to their past averaged
SFR are likely to deviate above the IRX--$\beta$ relation, i.e., have
larger $A_{\rm FUV}({\rm IRX})$ for a given $A_{\rm
  FUV}(\beta)$. Despite this qualitative agreement, the
\citeauthor{kong04} mean numerical relation for their UV-selected
sample of $50$ local starbursts is not a good fit to the FIR-selected
SFRS data as shown in Figure~\ref{fnuv-irx}. Regardless of numerical
values, all these studies agree that {\em galaxies with higher SFR
  are more obscured at fixed $\beta$} \citep[also
  see][]{Cortese2006,moore10,iglesias04}. At the low SFR end, galaxies with
$L_{60}<10^{9.3}$\,\Lsol, which at $z\approx0$ are mostly early-type
cluster galaxies, form two groups. Around 75 per cent of them are near the
mean IRX--$\beta$ relation, but the rest show $A_{\rm FUV}({\rm IRX})
\ll A_{\rm FUV}(\beta)$. {One possibility is that these galaxies
  have older stellar populations with intrinsically high values of
  $\beta$.} In the middle range 
$10^{9.3}<L_{60}<10^{10.7}$\,\Lsol, there is a general trend for
$A_{\rm FUV}({\rm IRX})$ to follow $A_{\rm FUV}(\beta)$ but with rms
scatter $\sim${0.34}\,dex. At $L_{60}>10^{10.7}$\,\Lsol, the scatter is
$\sim${0.56}\,dex. 

Some of the scatter in Figure~\ref{fnuv-irx} can be attributed to
intrinsic dispersion in the SFHs and metallicity of individual
galaxies {\citep{Cortese2006,kong04,johnson07,wilkins2012,Grasha2013}}. It is therefore
not surprising that the best-fit relation between $A_{\rm
  FUV}(\beta)$ and $A_{\rm FUV}({\rm IRX})$ matches well with the one
derived for a UV-selected sample of galaxies with $IRAS$ counterparts
\citep{seibert05}, similar to the SFRS sample used here, but is
significantly different from the one proposed by \citet{salim07} for
an optically-selected $z\sim0.1$ sample of galaxies. Using a sample
of galaxies from the SDSS and {\it GALEX}, \citet{treyer07} have also
confirmed that UV-based extinction corrections
\citep{seibert05,salim07,johnson07} over (under) estimate the
corrected UV luminosity for the lowest (highest) emission-line SFR
galaxies.  {In that context, it is remarkable that
  \citet{Casey2014} found qualitatively the same results we do despite
  having selected 5/6 of their sample galaxies in the UV.  (The other 1/6
of their sample was selected in the FIR, but most of those galaxies
have $\lfir > 10^{11}$\,\Lsol.) The similarity of our results implies
that they are not strongly biased by sample selection.}

\begin{figure}
\begin{center}
\includegraphics[width=0.48\textwidth,clip=true,bb=30 224 500 644]{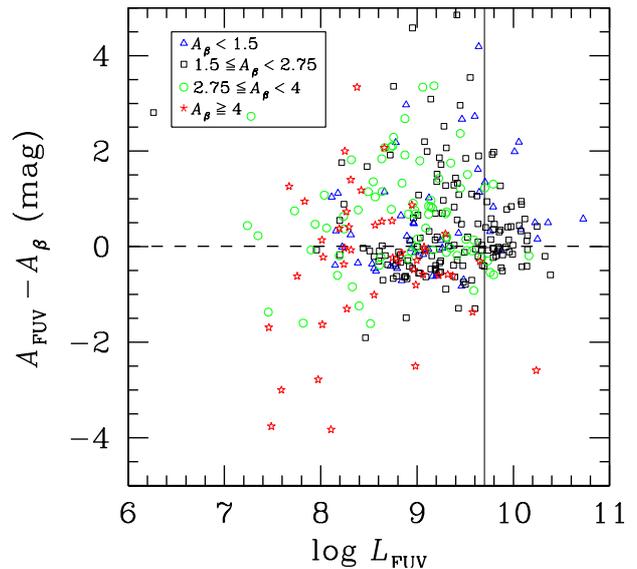}
\end{center}
\caption{Difference between bolometric extinction $A_{\rm FUV}({\rm
    IRX})$ and colour-excess extinction $A_{\rm FUV}(\beta)$ as a
  function of observed FUV luminosity. Point types indicate $A_{\rm
    FUV}(\beta)$ as indicated in the legend.  Dashed line indicates
  equality of the two extinction measures, i.e., for galaxies near
  the line, $A_{\rm FUV}(\beta)$ is a good predictor of $A_{\rm FUV}$
  and therefore of SFR. The solid line marks $L_{\rm FUV} =
  5\times10^9$\,\Lsol. }
\label{fig:lfuv}
\end{figure}

The geometry of dust and stars in a galaxy is a crucial element in
determining the attenuation at any given wavelength. However, while
the relation between FUV attenuation and IRX is almost independent of
dust geometry \citep{witt00}, the relation between $\beta$ and IRX
depends strongly on geometry of stars, gas, and dust {(probably
  the main effect according to \citealt{Casey2014})}, on dust grain
properties, and on dust clumpiness
\citep{witt00,charlot00,meurer99}\footnote{\citeauthor{witt00} and
  \citeauthor{charlot00} based their conclusions on IRX1600, which is
  the {\it{International Ultraviolet Explorer}} equivalent of IRX as
  defined above, but the same 
  results should hold for {\it GALEX} with $\lambda_{\rm
    eff}^{FUV}=1516$\,\AA.}. {For example, if young stars and
  dust are well mixed in an optically thick cloud, the extinction is
  high, but the UV light will emerge only from a layer near the
  surface and will show small $\beta$.  Small $\beta$ can also be
  seen if our line of sight happens to pass through a low-extinction
  `tunnel' to the young stars while their light emitted in other
  lines of sight is mostly absorbed.} These considerations probably
explain much of the scatter in using $\beta$ as an extinction
indicator (Figure~\ref{fnuv-irx}).  Figure~\ref{fig:lfuv} further
illustrates the problem of using UV data alone to estimate
SFR. Ignorance of the actual FIR emission gives a median (mean) error
in SFR of {0.22 (0.34)}\,dex and maximum error
{1.9}\,dex. There is no obvious way to know which galaxies will
have deviant SFRs, though there are some clues.  For 62 galaxies with
$L_{\rm FUV}>5\times10^9$\,\Lsol\ (and excluding two Seyfert galaxies
SFRS 263/270---\citealt{Marag2017b}), the median deviation is
{0.16\,dex}, and the maximum is {0.87\,dex}.  Similarly, for
{20} galaxies with {$\beta<-1.5$ (or equivalently $A_{\rm
    FUV}(\beta)<1.05$\,mag}, the median deviation is {0.15\,dex},
and only {one galaxy} deviates by more than {0.4}\,dex
{(though the deviation for that galaxy is 1.7\,dex)}.  Thus for
{$\sim$1/4} of the SFRS sample and presumably a larger fraction of
UV-selected samples, FIR estimates based on UV data are not
bad.  For the other {$\sim$3/4} of the galaxies, however, the
median deviation is {0.26}\,dex, and {28} per cent of the galaxies
show deviations $>$0.5\,dex.

In retrospect, it should not be surprising that $A_{\rm FUV}(\beta)$
gives a reasonable estimate of $A_{\rm FUV}$ when $A_{\rm
  FUV}(\beta)$ is small enough.  When $A_{\rm FUV} < 1$\,mag, of
order half or more of the UV light escapes, and the UV colour can
indicate extinction.  When extinction is larger, however, 
little  light from stars suffering high absorption escapes.
The light that does escape comes only from stars in
lines of sight that have low absorption, and only this low extinction
is measured.
Figure~11(a) of \citet{charlot00} illustrates the effect for a
simple mixed-slab model.  Real galaxies are even more complicated:
\citet{Goldader2002} showed that the emerging UV light is often
displaced from the main luminosity sources.  \new{Figure~\ref{image}
  shows some striking examples in the SFRS sample.}  Under these
conditions, the colour of the emerging light cannot indicate whether
there are many or few stars hidden by dust.

Another colour-excess extinction measure is the Balmer decrement, the
\hab\ flux ratio measured spectroscopically. \citet{Marag2017b} gave
nuclear Balmer decrements for the SFRS sample, most coming from SDSS
fiber spectra. However, \hab\ used here has an inherent bias because
the line ratio was measured from apertures 3\arcsec--3.5\arcsec\ in
diameter centred on each galaxy's nucleus.\footnote{{At the
    quartile and median distances of the SFRS galaxies, the 3\arcsec\
    SDSS fiber diameter corresponds to 370\,pc, 1.1\,kpc, and
    1.9\,kpc respectively.} \citeauthor{Marag2017b} gave long-slit
  Balmer decrements for 168 galaxies, but even those don't sample the
  entire galaxy disc. In order to compare all galaxies in our sample
  in a uniform way, we use here only the nuclear spectra even when
  long-slit spectra are available.} This introduces biases in two ways: \\
(i) galaxies must have significant Balmer emission and therefore high
nuclear star formation activity in order for emission lines to be
measureable, and \\
(ii) the circumnuclear regions of galaxies are dustier than the outer
disc \citep[\eg][]{popescu05,prescott07}, and hence for the nearby
galaxies in the SFRS sample the nuclear obscuration could exceed the
galaxy's average value. Therefore, the results are only a rough
indicator of the bolometric $A_{\rm FUV}$. A full analysis requires
optical spectra covering a larger fraction of the galaxies' areas
\citep[e.g.,][]{Cortese2006}.

\begin{figure}
\begin{center}
\includegraphics[width=0.48\textwidth,clip=true,bb=72 300 560 694]{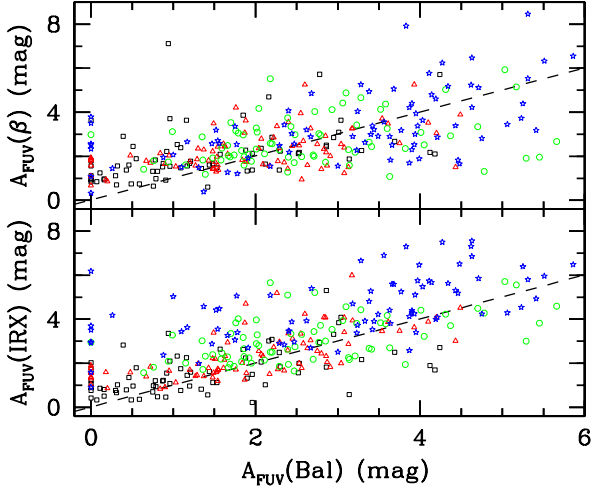}
\end{center}
\caption{Measures of FUV extinction as a function of \new{observed}
  Balmer decrement. {\it(Top)} FUV colour-excess extinction $A_{\rm
    FUV}(\beta)$ (Eq.~\ref{eq:beta}). {\it (bottom)} FUV bolometric
  extinction $A_{\rm FUV}({\rm IRX})$ (Eq.~\ref{afuv}). Abscissa
  comes from Eq.~\ref{eqn:abal}. Point types indicate quartiles of
  60\,\micron\ luminosity as indicated in the Figure~\ref{fusfr}
  legend, and the dashed lines mark equal values of extinction on
  both axes.}
\label{bd}
\end{figure}

To compute the FUV extinction implied by  the Balmer decrements,
we followed \citet{dominguez13}:
\begin{equation}
E(B-V)=1.97\, \rm log \bigg[\frac{(H\alpha/H\beta)_{obs}}{2.86}\bigg]
\label{eqn:bd}
\end{equation}
where (\hab)$_{\rm obs}$ is the observed Balmer
decrement.\footnote{Eq.~\ref{eqn:bd} assumes the \citet{calzetti00}
  reddening curves evaluated at the wavelengths of H$\alpha$ and
  H$\beta$ and the intrinsic, unreddened Balmer decrement $= 2.86$,
  appropriate for an electron temperature of $10^4$\,K and an
  electron density $n_e = 100$\,cm$^{-3}$ for Case~B recombination
  \citep{Hummer1987}.} The adopted reddening curve
\citep{calzetti00} translates the Equation~\ref{eqn:bd}
colour excess to
\begin{equation}
   A_{\rm FUV}({\rm Bal}) = 10.27 \cdot 0.44\cdot E(B-V) \quad,
\label{eqn:abal}
\end{equation}
where the factor 0.44 accounts for lower extinction to stars than to
ionized gas. Figure~\ref{bd} compares $A_{\rm FUV}(\rm Bal)$ with
other extinction measures for the SFRS galaxies.  $A_{\rm FUV}(\rm
Bal)$ is correlated with $A_{\rm FUV}({\rm IRX})$, but the scatter
is 0.60\,dex rms. Galaxies with $L_{60}>10^{10.5}$\,\Lsol\ have even
more scatter, 0.66\,dex. For galaxies with $L_{60}<10^{10.5}$\,\Lsol,
the scatter is 0.51\,dex, and $A_{\rm FUV}(\rm Bal)$ overestimates
$A_{\rm FUV}({\rm IRX})$ (and hence SFR) by {a median of 0.12\,dex}. 
{Using a  factor of 0.40 insted of 0.44 for the ratio of stellar to
  gas extinction increases the median overestimate to 0.17\,dex but
  decreases the rms scatter to 0.48\,dex.  A factor of 0.48 gives
  median overestimate 0.04\,dex but increases the scatter to
  0.55\,dex. These values represent a plausible range, but no fixed ratio
  will make the nuclear Balmer decrement a good predictor of $A_{\rm
    FUV}({\rm IRX})$.}
$A_{\rm FUV}(\beta)$ does a little better: for the whole sample, the
scatter between 
$A_{\rm FUV}(\beta)$ and $A_{\rm FUV}({\rm IRX})$ is {0.49}\,dex and
only {0.37}\,dex when $L_{60}<10^{10.5}$\,\Lsol. {Despite
  that, the averages of the estimates agree reasonably well:} $A_{\rm
  FUV}(\beta)$ 
{under}estimates $A_{\rm FUV}({\rm IRX})$ by 0.15\,dex for the
whole sample and {overestimates} by 0.03\,dex for
$L_{60}<10^{10.5}$\,\Lsol.  There 
is little correlation between $A_{\rm FUV}(\rm Bal)$ and $A_{\rm
  FUV}(\beta)$, which relation shows rms scatter 0.63\,dex.  These
results are in broad agreement with \citet{wijesinghe11}, who used
multi-wavelength data for a volume-limited sample of nearby galaxies
to show that there is a stronger correlation between $\beta$ and IRX
than between \hab\ and IRX but with a large scatter in both
(Figures~\ref{fnuv-irx} and \ref{bd}). Some of the scatter seen in
Figure~\ref{bd} may be attributed to the fact that $\beta$ depends
not only on the distribution of {dust and} young stars but also
on the age of 
that stellar population \citep{Grasha2013}, the contribution from older stellar
populations, metallicity, and the slope of the IMF.
 
The inability of reddening based on \hab\ to correct $L_{\rm
  FUV}$ in a way that determines FIR luminosity has previously been
reported by several authors \citep[\eg][]{wang96,buat99,buat02}. In
particular, \citet{buat02} showed that the correlation between dust
extinction and $L_{\rm FIR}$ is weak but gets worse for $B$-band,
H$\alpha$, or UV luminosities.  While colour-excess extinction
corrections may yield statistically useful SFRs for normal galaxies,
especially for low-dust samples selected at blue wavelengths,
identifying which galaxies are $L_{\rm FIR}\gtrsim10^{10.5}$\,\Lsol\
starburst galaxies requires data at longer wavelengths. 


\section{Comparison with previous results}
\label{discussion}

Using a heterogeneous sample of 249 galaxies from the literature,
\citet{bell03} presented a study similar to the present one. Our data
confirm Bell's principal conclusion: while FIR reliably represents
the star formation in ${\sim} L^*$ galaxies, it represents only a
fraction of it in lower-luminosity (${\sim}$0.01\,L$^*$)
galaxies. Another of \citet{bell03}'s suggestions was that radio
emission also underestimates the SFR in low-luminosity galaxies, and
that these corresponding underestimates are responsible for the
observed FIR--radio correlation. Figure~\ref{eta} confirms that
result and Bell's inference that the observed linear radio--FIR
correlation is a coincidence of both indicators underestimating the
SFR at low luminosities.  The remedy is a non-linear relation between
$L_{\rm 1.4 GHz}$ and SFR, as suggested by \citet{Davies2017} and
shown in Equation~\ref{radio-sfr2}.

\begin{figure} 
\begin{center}
\includegraphics[width=0.48\textwidth,clip=true,bb=72 290 600 700]{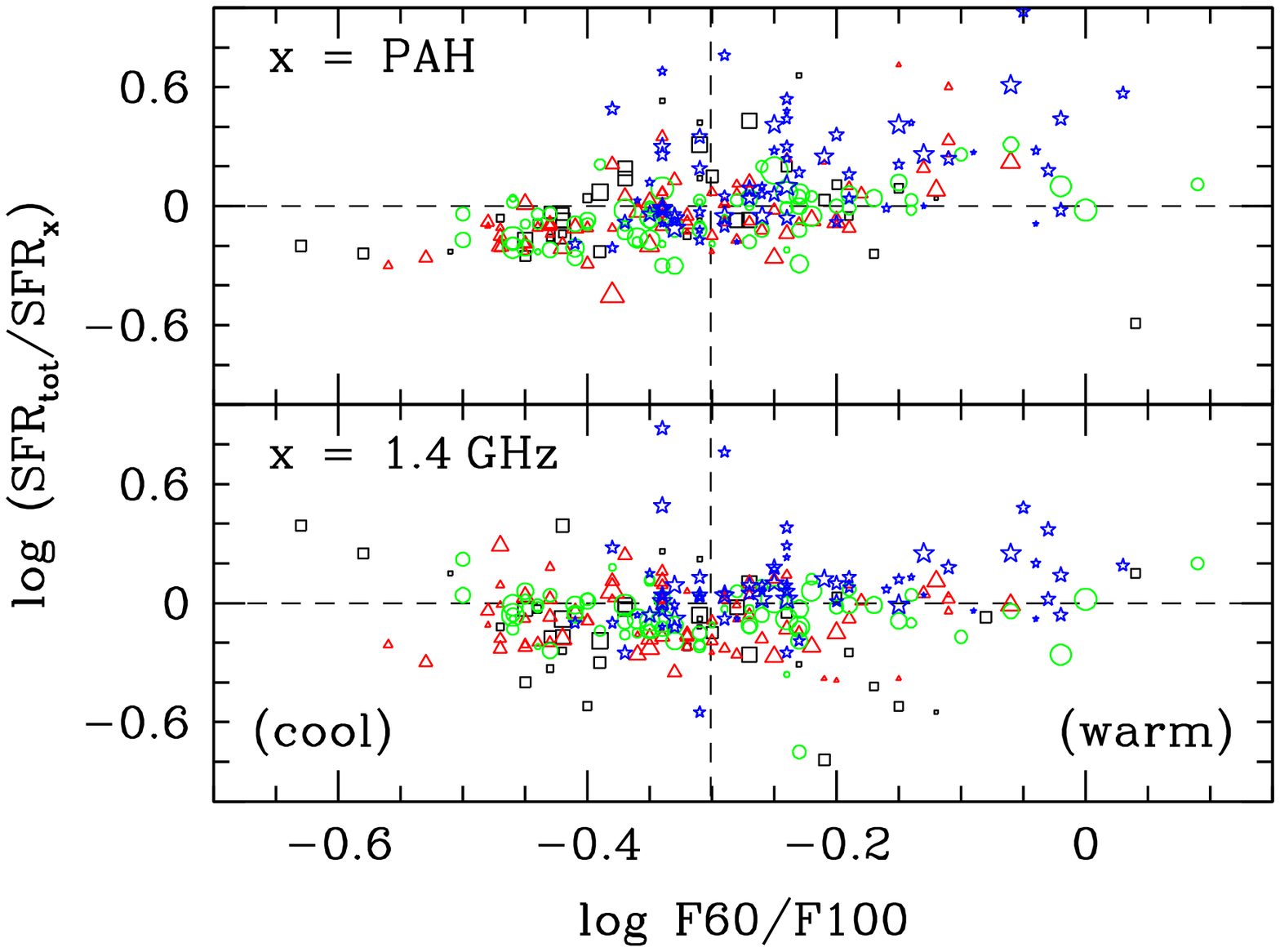}
\end{center}
\caption{\sfrfir\ relative to two other SFR measures as a function of
dust temperature.  The ratios on the ordinates are \sfrp\ (upper) and
non-linear \sfrrd\ (lower). The abscissa is the ratio of \IRAS\ 60 to
100~\micron\ flux densities. \new{Symbol} areas are proportional to
weights in the SFRS sample (Paper~I), and \new{galaxies} with $\rm
weight < 3$ (galaxies with \new{uncommon} properties) are not
shown. Point types indicate quartiles of 60\,\micron\ luminosity as
indicated in the Figure~\ref{fusfr} legend.  Horizontal dashed lines
show equality, and the vertical dashed line marks
$F_{60}/F_{100}=0.5$.}
\label{dtemp}
\end{figure}

\citet{bell03} also suggested a correlation between FIR dust
temperature $T_d$ and $\eta$. In their picture, hotter dust would imply
that active star formation is more important for dust heating, i.e.,
$\eta$ is smaller.  Indeed, galaxies with $F_{60}/F_{100} > 0.5$ are
generally termed starbursts \citep{rowan89}, and 146 of our sample  galaxies
fit this criterion. However, as shown in Figure~\ref{dtemp}, contrary
to the expectation, \sfrt/\sfrrd\ shows no correlation with $T_d$,
and \sfrt/\sfrp\ shows if anything an opposite 
correlation.  That is, if cool dust were being heated by an old
stellar population, \sfrt\ would over-estimate SFR, contrary to the
trend seen.

\begin{figure} 
\begin{center}
\includegraphics[width=0.48\textwidth,clip=true,bb=62 300 600 688]{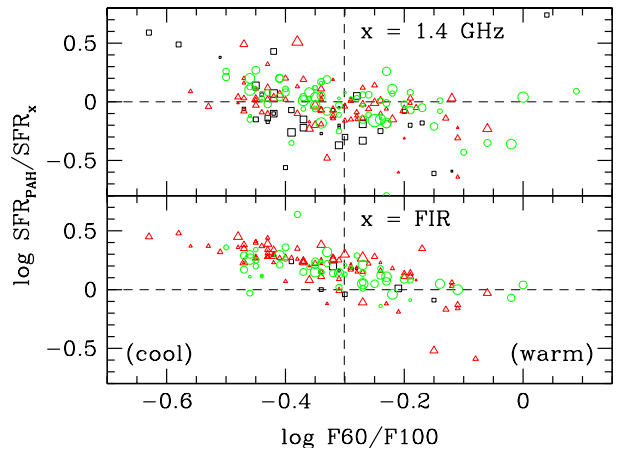}
\end{center}
\caption{Ratio of \sfrfir\ to \sfrp\ as a function of dust
  temperature.  The abscissa is the ratio of \IRAS\ 60 to
  100~\micron\ flux densities.  Point areas are proportional to
  weights in the SFRS sample (Paper~I), and points with $\rm weight <
  3$ are not shown. Point types indicate quartiles of 60\,\micron\
  luminosity as indicated in the Figure~\ref{fusfr} legend.
   The 
  horizontal dashed line shows equality, and the vertical dashed line
  marks $F_{60}/F_{100}=0.5$. Points in the highest quartile of
  $L_{60}$ are omitted \new{because 
    \sfrp\ has large scatter for them as shown in
    Figs.~\ref{frac-sfr} and~\ref{dtemp}}.}
\label{tdustp}
\end{figure}

The trend in \sfrt/\sfrp\ in Figure~\ref{dtemp} hints that $T_d$ may
have some relation to PAH emission.  That is confirmed by Figure
\ref{tdustp}, which shows that relative \sfrp\ is anti-correlated
with $T_d$.  
The strongest negative correlation is with \sfrfir, but {the
unweighted SFRS sample does not show a correlation of \sfrfir\ with
$T_d$ because the SFRS was constructed to cover the full range
of $T_d$ {\em uniformly} at each value of \lfir.  \sfrfir would
correlate with $T_d$ for the SFRS if the galaxies were properly
weighted.
Numerous studies have shown associations between high SFR, high sSFR,
high sSFR relative to the galaxy main sequence (`starburstiness'),
warm $T_d$, and low PAH emission
\citep[e.g.,][]{Elbaz2011,Nordon2012,Diaz2013,Stierwalt2014}. What
the SFRS sample shows is that {\em at fixed \lfir} (or SFR), higher
$T_d$ is associated with relatively smaller \lpah.}  We suggest that
warm dust is associated with a relative deficiency of
photo-dissociation regions, where the PAH emission
originates. {The most straightforward physical reason is high
  dust content in \hii\ regions of some galaxies.  Such dust grains,
  being relatively near the heating sources, would reach relatively
  high temperatures, and the energy they absorb could not escape to
  excite PAH molecules in surrounding PDRs. This mechanism was
  suggested \citep{Murata2014} to explain the PAH deficit in galaxies
  with high sSFR relative to their stellar masses.} Whether this
{explanation} is correct could {perhaps} be elucidated by
spatially resolved observations.

\citet{Davies2017} used galaxies from the GAMA survey to derive
conversions (linear and non-linear) from $L_{\rm 1.4\,GHz}$ to
SFR. These authors found a non-linear relation with slope 0.66 and
0.4\,dex scatter between $L_{\rm 1.4\,GHz}$ and their calculation of
\sfrt.  For the SFRS sample, the scatter between \sfrrd\ and our
\sfrt\ is only 0.25\,dex. The SFRS data prefer a steeper slope
$\sim$0.72 of the SFR--$L_{\rm 1.4\,GHz}$ relation, nearly equal to
the slope of 0.75 found by \citeauthor{Davies2017} using the MAGPHYS
\citep{daCunha2008} estimate of SFR instead of their \sfrt.
  
\begin{figure}
\begin{center}
\includegraphics[width=0.48\textwidth,clip=true,bb=72 300 600 688]{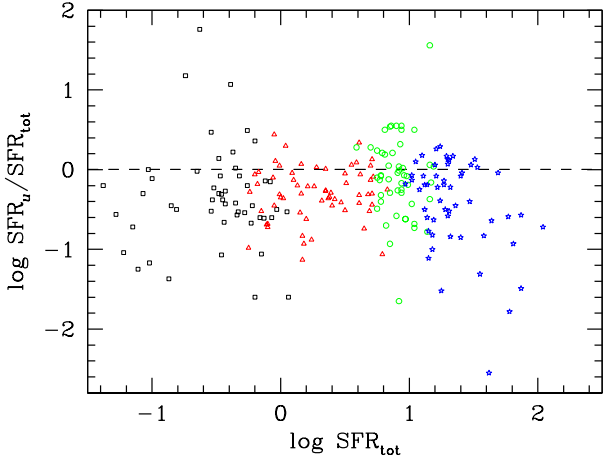}
\end{center}
\caption{Ratio of $u$-band SFR \citep[][Eq.~B8]{hopkins03} to
  \sfrt. Point types indicate quartiles of 60\,\micron\ luminosity as
  indicated in the Figure~\ref{fusfr} legend, and the dashed line
  shows equality between the two measures. SFRS~17 (=NGC~2552) is
  omitted because it is very faint at $u=22.581$\,mag.}
\label{sfru}
\end{figure}  
  
Another common SFR indicator is the ground-based $u$-band (${\sim}
3500$\,\AA\ in SDSS and similar data sets)
luminosity. \citet{hopkins03} used a sample of 3079 galaxies observed
at 1.4\,GHz by FIRST\footnote{Faint Images of the Radio Sky at Twenty
  cm---\citet{White1997}} and by SDSS to compare SFR indicators
based on H$\alpha$, [\ion{O}{ii}], $u$-band (here denoted \sfru), and
FIR luminosities against \sfrr.  A critical element of the
\citeauthor{hopkins03} SFR calculation (their Eq.~B8) was the
$u$-band extinction correction, which they derived from each galaxy's
Balmer decrement. \citet{davies16} used the same \sfru\ metric (among
12 that they examined) in a sample of morphologically selected spiral
galaxies ($0<z<0.13$) from the Galaxy and Mass Assembly (GAMA)
survey \citep{Driver2016}.  One key difference is that
\citeauthor{davies16} derived the 
$u$-band extinction from fitting each galaxy's spectral energy
distribution, and they also put in a correction, based on $u-g$
colour, for $u$-band radiation from older stars.  Figure~\ref{sfru}
compares \sfru\ to \sfrt\ for 218 SFRS galaxies where SDSS data are
available.  The derived values of \sfru\ are a
factor of two lower than \sfrt\ on average, and the rms scatter is
0.6~dex. This applies even for low-luminosity galaxies, despite
\sfrfuv\ often being dominant over \sfrfir\
(Fig.~\ref{fusfr}). Correcting for $u$-band emission by old stars
\citep[][Eq.~12]{davies16} would make the discrepancy worse,
increasing the median \sfru\ deficit to 0.5~dex and the scatter to
0.67~dex. The systematic errors and large scatter make $u$-band luminosity an
unreliable SFR indicator for galaxies with dust such as the SFRS sample.

\citet{bell03} showed that the H$\alpha$ and FUV extinctions loosely
correlate with each other, with the former being around half of the
latter.  Our observations (Figure~\ref{bd}) agree with the
correlation, but neither indicator is a reliable measure of
bolometric extinction (Figure~\ref{fnuv-irx}).
\citet{wang16} similarly studied a subsample of 745 galaxies from the
GAMA and \Her\ Astrophysical Terahertz Large Area
Survey ({H-ATLAS}) to test correlations between multi-wavelength SFR
tracers. \citet{wang16} derived SFRs using FIR, sub-mm,
dust-corrected UV photometric data, and H$\alpha$ emission line
luminosities, the last two corrected for attenuation using the Balmer
decrement. \citeauthor{wang16} found that UV data can be reconciled
with the attenuation-corrected H$\alpha$ SFR and their version of
\sfrt\ (computed assuming $\eta=0.54$; Eq.~\ref{tot-sfr}) after
applying an attenuation correction based on IRX. In agreement with our
results, \citeauthor{wang16} showed that the UV spectral slope
$\beta$ is not a reliable attenuation indicator on its
own. \citeauthor{wang16} also found that the attenuation correction
factor depends on stellar mass, redshift, and dust temperature but is
independent of the H$\alpha$ equivalent width and S\'{e}rsic
index. To summarize, none of the colour-excess indicators we
have tested can be considered reliable for general galaxy samples.

\section{Summary and prospects}
\label{conclusions}

The main conclusion of our work is that for local galaxies, global
SFRs can be derived consistently from radio continuum, FIR+UV, or
8.0\,\micron~PAH emission with scatter ${\la}0.3$\,dex in SFR over four
orders of magnitude in galaxy luminosity. In particular,
{the SFRS results confirm and quantify}:

\begin{itemize}
\item For measuring SFR from 1.4\,GHz radio observations, the
  preferred calibration is non-linear with a slope near the 0.75 value
  found also by \citet{Davies2017}.
\item The distributions of \sfrt, non-linear \sfrrd, and \sfrp\ show
  similar statistical properties. {We have presented mutually
    consistent (within 5 per cent) calibrations for these measures of SFR.}
\item \lfir\ captures most of the emergent
  luminosity for most luminous ($\log L_{60}\geq10.5$\,\Lsol) galaxies
  and is therefore a good measure of their SFR. Lower-luminosity
  galaxies tend to have more of their emission in the UV, which is
  therefore needed to estimate their SFR.  Different numerical
  prescriptions, such as adding \sfrfuv\ or \sfrnuv\ or using a
  bolometric extinction correction for the UV light, give
  statistically similar results.
\item SFR estimates obtained from UV data alone are subject to large
  uncertainties in the extinction corrections.  At fixed UV colour or
  spectral slope $\beta$, galaxies with $L_{\rm FUV}<5\times
  10^9$\,\Lsol\ show a broad range in $A_{\rm FUV}$. Therefore the UV
  spectral slope is not a good measure of the correction needed, and
  extinction corrections based on UV colour may yield \sfrfuv\
  differing from the total SFR by up to 2\,dex. {In contrast, for
    galaxies with $L_{\rm FUV}>5\times 10^9$\,\Lsol\ or $\beta<
    -1.5$, $L_{\rm FUV}$ corrected by extinction based on UV spectral
    slope ($A_{\rm FUV}(\beta)$) can measure SFR with rms scatter
    $\sim$0.24\,dex.}
\end{itemize}
{The SFRS data also reveal:}
\begin{itemize}
\item For galaxies with  ${\rm FUV}-[3.6]\la2$, PAH luminosity
  underestimates SFR by up to $\sim$1\,dex.
\item The FIR-selected SFRS sample shows a {surprising}
  preference for $\eta=0$ for obtaining \sfrt. In other words,
  {when using \lfir\ to deduce SFR,} accounting for dust heating
  by an {older} stellar population unrelated to current star
  formation is unimportant.  The existence of the star-forming-galaxy
  main sequence and the sub-galactic main sequence \citep[][and
  references therein]{Marag2017a} suggests that at least part of the
  explanation is the close association between current star formation
  and the pre-existing stellar population. {Much larger samples
    or better theoretical understanding of the SFR tracers
    will be needed for accurate measurements of $\eta$.}
\item Dust temperature does not correlate with most measures of SFR,
  but there is a close relation between dust temperature and
  \sfrp/\sfrfir.  This needs to be explored in spatially resolved
  galaxies.
\end{itemize}

{Consistency of various SFR indicators does not prove they are
  correct.} 
Some authors \citep[e.g.,][]{Boquien2014,daSilva2014} have suggested
that bursty star formation histories can {cause}  measured SFRs
{to deviate from the true values}.  We
have not examined that suggestion because the purpose of this paper
is to {inter-}compare empirical SFR indicators, but future work should
investigate this possibility. Future work should also include better
decomposition of the SFRS galaxies into AGN and star-forming
components and the correlations of SFR indicators with each component.
Additional SFR indicators such as H$\alpha$ line flux, full SED
fitting with derivation of reddening and corrected UV flux, and {\it
  Spitzer}/MIPS 24\,\micron\ or {\it WISE} 25\,\micron\ should also be
examined.

\section*{Acknowledgements}

This work is based in part on data obtained with the {\it Spitzer Space
Telescope}, which is operated by the Jet Propulsion Laboratory,
California Institute of Technology under a contract with
NASA. Support for this work was provided by NASA.
 This research has made use of the NASA/IPAC Extragalactic Database (NED), which is operated by the Jet Propulsion Laboratory, 
 California Institute of Technology, under contract with the National Aeronautics and Space Administration.

Funding for the Sloan Digital Sky Survey IV has been provided by
the Alfred P. Sloan Foundation, the U.S. Department of Energy Office of
Science, and the Participating Institutions. SDSS-IV acknowledges
support and resources from the Center for High-Performance Computing at
the University of Utah. The SDSS web site is \url{www.sdss.org}.
SDSS-IV is managed by the Astrophysical Research Consortium for the
Participating Institutions of the SDSS Collaboration including the
Brazilian Participation Group, the Carnegie Institution for Science,
Carnegie Mellon University, the Chilean Participation Group, the
French Participation Group, Harvard-Smithsonian Center for
Astrophysics, Instituto de Astrof\'isica de Canarias, The Johns
Hopkins University, Kavli Institute for the Physics and Mathematics
of the Universe (IPMU) / University of Tokyo, Lawrence Berkeley
National Laboratory, Leibniz Institut f\"ur Astrophysik Potsdam
(AIP), Max-Planck-Institut f\"ur Astronomie (MPIA Heidelberg),
Max-Planck-Institut f\"ur Astrophysik (MPA Garching),
Max-Planck-Institut f\"ur Extraterrestrische Physik (MPE), National
Astronomical Observatory of China, New Mexico State University, New
York University, University of Notre Dame, Observat\'ario Nacional /
MCTI, The Ohio State University, Pennsylvania State University,
Shanghai Astronomical Observatory, United Kingdom Participation
Group, Universidad Nacional Aut\'onoma de M\'exico, University of
Arizona, University of Colorado Boulder, University of Oxford,
University of Portsmouth, University of Utah, University of Virginia,
University of Washington, University of Wisconsin, Vanderbilt
University, and Yale University.

This publication makes use of data products from the Two Micron All
Sky Survey, which is a joint project of the University of
Massachusetts and the Infrared Processing and Analysis
Center/California Institute of Technology, funded by the National
Aeronautics and Space Administration and the National Science
Foundation.

Mahajan gratefully acknowledges support from Smithsonian Institution
Endowment Grant for the SAO Predoctoral Fellowship which helped lay
the foundation of this work. Mahajan is funded by the INSPIRE Faculty
award (DST/INSPIRE/04/2015/002311), Department of Science and
Technology (DST), Government of India. Barmby acknowledges support
from an NSERC Discovery Grant. 
Maragkoudakis acknowledges funding from the European Research Council
under the European Union's Seventh 
Framework Programme (FP/2007-2013)/ERC Grant Agreement
number 617001. This project has received funding from the European
Union's Horizon 2020 research and innovation program under the Marie
Sklodowska-Curie RISE action, grant agreement number 691164 (ASTROSTAT).






\clearpage
\appendix
\onecolumn
\renewcommand{\thetable}{A\arabic{table}}
\setlength{\LTleft}{36pt}
\setlength{\LTright}{0pt}

\begin{center}
\begin{longtable}{clrccccc}
\kill
{\footnotesize}
\null\\
\caption{{\it GALEX} data for SFRS galaxies} \\
\hline\hline
SFRS$^a$ & 
Name & 
{$D$(Mpc)$^a$} &
FUV$^b$ & 
$\Delta$FUV & 
NUV$^b$ & 
$\Delta$NUV
& $E(B-V)$$^c$  
\\
\hline\endhead
\hline\endfoot
1&IC 486&114.4&18.179&0.029&17.575&0.014&0.040 \\
2&IC 2217&76.1&16.434&0.013&15.939&0.006&0.041 \\
3&NGC 2500&15.0&13.925&0.004&13.785&0.002&0.040 \\
5&MCG 6-18-009&164.4&17.890&0.026&17.064&0.011&0.052 \\
8&NGC 2532&77.6&15.417&0.008&14.862&0.004&0.054 \\
9&UGC 4261&93.2&16.485&0.014&16.159&0.007&0.055 \\
10&NGC 2535&61.6&15.708&0.010&15.290&0.004&0.043 \\
11&NGC 2543&26.3&15.986&0.011&15.516&0.005&0.069 \\
12&NGC 2537&15.0&14.964&0.007&14.752&0.003&0.054 \\
13&IC 2233&13.7&15.000&0.007&14.805&0.004&0.052 \\
14&IC 2239&88.5&19.177&0.046&18.041&0.016&0.053 \\
15&UGC 4286&75.1&17.372&0.020&16.945&0.009&0.047 \\
16&UGC 4306&36.0&18.679&0.036&18.131&0.019&0.069 \\
17&NGC 2552&11.4&14.736&0.006&14.587&0.003&0.047 \\
18&IC 2339&79.3&17.100&0.018&16.702&0.009&0.047 \\
19&IRAS 08234+1054&272.6&19.653&0.055&18.961&0.028&0.050 \\
20&IRAS 08269+1514&134.5&20.736&0.074&19.758&0.028&0.033 \\
21&NGC 2604&36.3&15.008&0.007&14.506&0.003&0.041 \\
22&NGC 2608&36.3&16.004&0.011&15.414&0.006&0.039 \\
24&NGC 2623&81.6&17.512&0.022&16.861&0.009&0.041 \\
25&CGCG 120-018&107.9&20.649&0.078&19.397&0.026&0.034 \\
29&IRAS 08512+2727&265.3&18.413&0.032&17.805&0.014&0.034 \\
31&IRAS 08538+4256&121.2&19.898&0.062&18.661&0.022&0.024 \\
32&IRAS 08550+3908&367.8&19.943&0.063&19.615&0.032&0.029 \\
33&NGC 2718&57.4&15.772&0.010&15.298&0.005&0.071 \\
34&NGC 2712&30.9&15.372&0.008&15.045&0.004&0.020 \\
35&NGC 2719&51.1&16.235&0.012&15.759&0.005&0.033 \\
36&IRAS 08572+3915NW&244.3&20.653&0.054&20.212&0.026&0.027 \\
37&IRAS 08579+3447&273.5&19.027&0.042&18.260&0.017&0.028 \\
38&NGC 2731&35.0&16.621&0.014&16.099&0.007&0.065 \\
39&NGC 2730&58.9&15.753&0.010&15.390&0.005&0.027 \\
40&IC 2431&209.0&18.787&0.038&18.173&0.017&0.050 \\
41&NGC 2750&37.0&14.993&0.007&14.637&0.004&0.036 \\
42&IC 2434&104.5&17.590&0.022&17.041&0.010&0.020 \\
43&NGC 2761&125.0&18.753&0.037&17.851&0.014&0.036 \\
44&NGC 2773&80.4&18.353&0.031&17.646&0.013&0.048 \\
45&NGC 2776&36.0&14.520&0.006&14.089&0.003&0.014 \\
46&NGC 2789&93.6&18.281&0.031&17.218&0.011&0.026 \\
48&NGC 2824&42.5&19.236&0.047&17.950&0.015&0.032 \\
49&IRAS 09184+4356&170.1&20.281&0.073&19.564&0.032&0.015 \\
50&CGCG 238-041&131.5&18.122&0.022&17.627&0.010&0.015 \\
51&UGC 4985&143.4&19.735&0.057&19.442&0.033&0.036 \\
52&NGC 2854&25.0&16.742&0.015&16.210&0.007&0.018 \\
53&UGC 5046&64.9&17.862&0.025&17.020&0.010&0.034 \\
54&UGC 5055&110.6&16.568&0.014&16.112&0.007&0.034 \\
55&NGC 2893&24.0&16.229&0.012&15.674&0.005&0.023 \\
56&MCG 3-24-062&66.3&18.404&0.032&18.130&0.017&0.045 \\
57&CGCG 238-066&147.0&19.008&0.042&18.281&0.018&0.012 \\
58&UGC 5097&72.5&16.210&0.012&15.841&0.006&0.039 \\
59&CGCG 289-012&172.5&17.394&0.020&16.948&0.009&0.020 \\
60&MCG 8-18-013&110.9&18.171&0.029&17.572&0.013&0.024 \\
61&CGCG 181-068&100.6&20.206&0.070&19.396&0.032&0.014 \\
62&NGC 2936&100.5&16.655&0.015&16.210&0.007&0.034 \\
63&NGC 2955&103.5&16.395&0.013&15.853&0.006&0.011 \\
64&CGCG 182-010&175.1&18.727&0.037&18.102&0.016&0.012 \\
65&UGC 5228&28.2&16.937&0.017&16.440&0.009&0.164 \\
67&NGC 3015&108.8&17.958&0.026&17.232&0.011&0.084 \\
68&MCG 2-25-039&77.6&18.050&0.027&17.468&0.012&0.031 \\
69&NGC 3020&18.3&14.492&0.005&14.200&0.003&0.038 \\
70&NGC 3049&18.3&15.477&0.009&15.122&0.004&0.038 \\
71&NGC 3055&25.0&15.203&0.008&14.744&0.004&0.052 \\
72&IC 2520&26.4&16.496&0.014&16.171&0.007&0.021 \\
73&UGC 5403&33.0&18.118&0.028&17.502&0.012&0.031 \\
74&UGC 5459&25.8&15.529&0.009&14.508&0.003&0.007 \\
75&MCG 5-24-022&92.1&19.488&0.150&18.203&0.016&0.022 \\
76&IC 2551&94.9&17.990&0.027&17.004&0.010&0.036 \\
77&IRAS 10106+2745&215.6&21.136&0.087&19.952&0.030&0.054 \\
78&NGC 3162&26.4&14.479&0.005&14.129&0.003&0.023 \\
79&IRAS 10120+1653&517.2&23.088&0.158&21.807&0.069&0.036 \\
80&NGC 3190&26.4&17.508&0.022&16.277&0.007&0.025 \\
81&IC 602&57.6&15.492&0.009&15.184&0.004&0.026 \\
82&NGC 3191&134.0&16.243&0.012&15.834&0.006&0.012 \\
83&NGC 3206&25.8&14.391&0.005&14.098&0.003&0.014 \\
84&UGC 5613&139.8&17.585&0.022&16.605&0.008&0.014 \\
85&UGC 5644&137.6&17.328&0.020&16.889&0.010&0.042 \\
86&NGC 3245&20.9&17.745&0.024&16.343&0.009&0.025 \\
88&MCG 7-22-012&66.0&18.827&0.039&18.130&0.016&0.011 \\
89&IRAS 10276+1119&271.3&18.174&0.029&17.905&0.015&0.033 \\
90&NGC 3265&21.8&17.273&0.019&16.784&0.009&0.024 \\
91&UGC 5713&95.0&18.375&0.032&18.009&0.018&0.019 \\
93&UGC 5720&20.0&15.071&0.007&14.831&0.004&0.012 \\
92&NGC 3274&10.0&14.687&0.006&14.479&0.003&0.024 \\
95&NGC 3306&46.6&16.448&0.013&16.073&0.006&0.025 \\
96&NGC 3323&79.5&16.081&0.011&15.677&0.005&0.024 \\
97&IC 2598&89.1&18.408&0.032&17.619&0.013&0.029 \\
98&NGC 3338&21.4&14.300&0.005&13.847&0.003&0.031 \\
99&NGC 3353&16.0&14.868&0.006&14.667&0.003&0.007 \\
100&UGC 5881&93.0&18.495&0.033&17.688&0.014&0.030 \\
101&NGC 3370&20.9&14.845&0.006&14.441&0.003&0.031 \\
102&NGC 3381&25.7&15.116&0.007&14.728&0.004&0.020 \\
103&UGC 5941&107.0&19.109&0.044&17.918&0.015&0.012 \\
104&NGC 3413&16.2&15.493&0.009&15.109&0.004&0.023 \\
105&NGC 3408&138.0&16.989&0.017&16.448&0.007&0.012 \\
106&NGC 3430&28.4&20.146&0.069&19.361&0.028&0.024 \\
107&CGCG 95-055&25.7&14.602&0.006&14.168&0.003&0.031 \\
109&UGC 6074&38.0&18.289&0.031&17.441&0.012&0.015 \\
110&NGC 3495&17.5&15.291&0.008&14.726&0.004&0.046 \\
111&UGC 6103&91.7&16.208&0.012&15.676&0.005&0.009 \\
112&MCG 7-23-019&150.6&17.344&0.020&16.920&0.009&0.009 \\
113&UGC 6135&90.9&16.551&0.014&15.941&0.006&0.008 \\
114&CGCG 241-078&110.9&20.440&0.215&19.183&0.025&0.009 \\
115&IRAS 11069+2711&296.4&19.304&0.036&19.241&0.021&0.021 \\
116&IC 676&26.9&17.796&0.025&16.839&0.009&0.025 \\
117&IRAS 11102+3026&129.6&22.427&0.135&20.895&0.042&0.021 \\
118&IC 2637&128.2&17.409&0.021&16.695&0.008&0.022 \\
119&MCG 9-19-013&201.8&19.958&0.063&19.265&0.028&0.016 \\
120&7Zw 384&340.4&18.649&0.028&18.004&0.012&0.008 \\
121&IRAS 11167+5351&447.3&20.357&0.053&20.003&0.027&0.014 \\
122&NGC 3633&30.0&18.619&0.036&17.779&0.014&0.062 \\
124&NGC 3656&37.0&17.272&0.020&16.487&0.008&0.015 \\
126&NGC 3664&26.9&14.685&0.006&14.441&0.003&0.045 \\
127&NGC 3666&16.3&15.532&0.009&15.020&0.004&0.033 \\
128&IC 691&16.0&16.560&0.014&16.251&0.007&0.014 \\
129&NGC 3686&21.0&14.432&0.005&13.890&0.003&0.024 \\
130&UGC 6469&102.6&16.334&0.013&15.909&0.006&0.037 \\
131&NGC 3690&52.6&15.996&0.008&15.511&0.004&0.017 \\
132&IC 698&96.8&18.445&0.033&17.421&0.013&0.043 \\
133&IRAS 11267+1558&736.6&22.046&0.090&21.665&0.045&0.045 \\
134&NGC 3705&16.3&15.009&0.007&14.517&0.003&0.046 \\
135&MCG 3-29-061&67.5&18.785&0.038&18.054&0.016&0.022 \\
136&NGC 3720&89.8&16.934&0.012&15.922&0.005&0.029 \\
137&NGC 3729&17.1&15.626&0.009&15.068&0.004&0.011 \\
139&NGC 3758&131.6&17.700&0.023&16.834&0.020&0.025 \\
140&UGC 6583&93.2&16.851&0.016&16.291&0.007&0.026 \\
142&NGC 3769&17.1&15.163&0.007&14.752&0.003&0.023 \\
143&NGC 3773&16.3&15.240&0.008&14.963&0.004&0.027 \\
144&NGC 3781&103.5&20.374&0.076&18.764&0.021&0.026 \\
145&UGC 6625&158.2&16.260&0.012&15.778&0.005&0.026 \\
146&NGC 3808&107.2&18.376&0.032&17.644&0.013&0.026 \\
148&NGC 3822&94.6&17.915&0.026&16.914&0.009&0.056 \\
149&UGC 6665&85.0&15.735&0.010&15.322&0.004&0.025 \\
150&MCG 3-30-051&90.4&17.574&0.022&17.173&0.010&0.020 \\
152&UGC 6732&53.6&19.415&0.050&17.479&0.012&0.020 \\
153&IC 730&93.1&18.778&0.038&18.039&0.015&0.021 \\
155&NGC 3912&22.5&16.286&0.012&15.621&0.005&0.021 \\
156&NGC 3928&16.9&15.658&0.009&15.220&0.004&0.020 \\
157&NGC 3934&61.6&19.687&0.059&18.745&0.023&0.055 \\
158&UGC 6865&91.2&18.342&0.031&17.340&0.011&0.020 \\
159&UGC 6901&107.6&18.684&0.036&17.456&0.012&0.015 \\
160&CGCG 013-010&172.3&19.818&0.060&18.646&0.023&0.016 \\
161&NGC 3991&55.6&14.581&0.006&14.449&0.003&0.022 \\
162&NGC 4004&57.9&15.813&0.010&15.454&0.005&0.025 \\
163&NGC 4014&62.6&17.129&0.018&16.448&0.008&0.065 \\
164&NGC 4010&17.1&16.142&0.012&15.608&0.005&0.025 \\
165&NGC 4018&72.6&18.075&0.028&17.290&0.011&0.021 \\
166&NGC 4020&14.3&15.458&0.009&15.122&0.004&0.017 \\
167&IRAS 11571+3003&218.6&21.136&0.095&20.257&0.038&0.018 \\
169&UGC 7016&110.3&18.982&0.042&17.906&0.016&0.038 \\
168&UGC 7017&55.3&16.731&0.015&16.387&0.007&0.019 \\
170&MCG 3-31-030&13.1&16.820&0.016&16.406&0.007&0.030 \\
171&NGC 4062&16.3&14.937&0.007&14.360&0.003&0.025 \\
172&NGC 4064&8.5&17.034&0.018&15.889&0.006&0.021 \\
173&CGCG 098-059&102.3&19.226&0.046&18.020&0.017&0.049 \\
174&NGC 4116&16.0&14.278&0.005&14.068&0.003&0.022 \\
175&NGC 4136&16.3&14.049&0.005&13.741&0.003&0.018 \\
176&NGC 4150&13.7&17.726&0.024&16.326&0.007&0.018 \\
177&IRAS 12086+1441&13.1&21.100&0.068&20.507&0.032&0.036 \\
178&NGC 4162&42.5&15.549&0.009&14.938&0.004&0.035 \\
179&NGC 4178&16.8&14.115&0.005&13.734&0.002&0.028 \\
181&NGC 4189&16.8&15.325&0.008&14.798&0.004&0.033 \\
183&NGC 4204&10.0&15.188&0.008&14.946&0.004&0.034 \\
184&NGC 4207&16.8&17.284&0.019&16.604&0.008&0.017 \\
185&UGC 7286&115.4&19.081&0.043&18.184&0.016&0.020 \\
186&NGC 4234&30.0&15.426&0.008&15.011&0.004&0.018 \\
187&NGC 4237&16.8&16.249&0.012&15.316&0.005&0.030 \\
188&NGC 4244&4.3&13.045&0.003&12.546&0.001&0.021 \\
189&NGC 4253&64.9&17.589&0.022&16.839&0.009&0.020 \\
190&MCG 3-32-005&13.1&18.598&0.035&17.720&0.013&0.028 \\
191&NGC 4290&37.0&15.859&0.027&15.346&0.005&0.014 \\
192&NGC 4294&16.8&14.544&0.006&14.165&0.003&0.034 \\
193&NGC 4314&16.3&15.848&0.010&14.951&0.004&0.025 \\
194&NGC 4385&29.0&15.574&0.009&15.213&0.004&0.025 \\
196&NGC 4396&16.8&15.137&0.007&14.790&0.004&0.026 \\
197&NGC 4412&30.6&15.367&0.008&14.971&0.004&0.018 \\
199&NGC 4420&17.6&15.005&0.007&14.484&0.003&0.018 \\
200&NGC 4424&6.0&16.296&0.013&15.418&0.005&0.021 \\
201&NGC 4435&16.7&18.515&0.035&16.763&0.009&0.030 \\
202&NGC 4438&16.8&15.844&0.010&15.241&0.004&0.028 \\
203&NGC 4448&16.3&16.310&0.013&15.585&0.006&0.024 \\
205&NGC 4470&16.8&15.238&0.008&14.868&0.004&0.024 \\
207&NGC 4491&16.8&17.596&0.023&16.903&0.009&0.042 \\
209&NGC 4495&74.2&17.244&0.019&16.646&0.008&0.021 \\
210&IC 3476&16.8&15.077&0.007&14.497&0.003&0.036 \\
211&NGC 4509&11.1&15.412&0.008&15.238&0.004&0.012 \\
212&NGC 4519&16.8&14.309&0.005&13.955&0.002&0.020 \\
213&NGC 4548&16.2&14.973&0.007&14.462&0.003&0.038 \\
214&IRAS 12337+5044&172.5&18.230&0.073&17.693&0.013&0.012 \\
216&NGC 4592&11.1&14.177&0.005&13.840&0.003&0.022 \\
217&NGC 4607&16.8&17.943&0.027&17.202&0.011&0.032 \\
218&NGC 4625&9.2&15.283&0.008&14.986&0.004&0.018 \\
219&NGC 4630&15.6&15.493&0.009&15.123&0.004&0.030 \\
222&MCG 5-30-069&74.0&20.277&0.073&18.822&0.022&0.015 \\
223&IC 3721&98.5&17.723&0.024&17.014&0.010&0.019 \\
224&NGC 4670&14.3&14.313&0.005&14.115&0.003&0.015 \\
226&MCG 7-26-051&146.9&19.704&0.057&18.889&0.024&0.022 \\
228&NGC 4688&15.6&14.740&0.006&14.625&0.004&0.031 \\
229&NGC 4704&122.8&16.979&0.017&16.555&0.008&0.017 \\
231&IRAS 12468+3436&498.3&24.890&0.281&22.798&0.075&0.014 \\
233&MCG 8-23-097&131.2&19.748&0.058&18.866&0.023&0.013 \\
234&NGC 4747&14.3&16.444&0.013&16.179&0.008&0.010 \\
235&UGC 8017&107.1&17.548&0.022&17.087&0.010&0.010 \\
236&NGC 4765&15.6&15.199&0.008&14.776&0.004&0.040 \\
237&VCC2096&13.1&18.511&0.033&17.828&0.014&0.051 \\
238&UGC 8041&23.0&14.962&0.007&14.636&0.003&0.022 \\
240&NGC 4837&132.5&16.658&0.015&16.204&0.007&0.014 \\
241&UM530&282.7&18.851&0.027&18.506&0.013&0.021 \\
242&NGC 4861&18.5&14.628&0.006&14.430&0.003&0.010 \\
244&NGC 4922&107.2&19.225&0.046&18.509&0.019&0.011 \\
245&UGC 8179&222.1&17.509&0.022&17.152&0.011&0.013 \\
246&NGC 5001&134.8&18.223&0.030&17.100&0.010&0.016 \\
247&IC 856&64.3&17.795&0.024&17.235&0.010&0.035 \\
248&UGC 8269&124.1&20.814&0.091&19.745&0.034&0.012 \\
250&NGC 5012&40.2&15.090&0.007&14.512&0.003&0.014 \\
249&NGC 5014&18.5&16.779&0.015&16.143&0.007&0.008 \\
251&IRAS 13116+4508&258.3&22.285&0.126&21.000&0.047&0.016 \\
252&IC 860&54.5&20.272&0.306&18.316&0.019&0.013 \\
253&IRAS 13144+4508&381.8&18.892&0.029&18.605&0.015&0.018 \\
254&NGC 5060&97.4&17.122&0.018&16.433&0.007&0.034 \\
255&UGC 8357&146.9&17.289&0.019&16.680&0.008&0.026 \\
256&UGC 8361&106.3&18.921&0.040&18.137&0.016&0.038 \\
257&IC 883&104.7&17.703&0.024&16.939&0.010&0.013 \\
258&NGC 5100&142.2&18.101&0.028&17.289&0.011&0.029 \\
259&NGC 5104&87.8&18.082&0.028&17.534&0.013&0.023 \\
260&NGC 5107&18.5&15.420&0.008&15.063&0.004&0.011 \\
262&NGC 5123&123.4&16.085&0.011&15.604&0.005&0.013 \\
263&IRAS 13218+0552&850.44&20.564&0.049&21.164&0.037&0.031 \\
264&IRAS 13232+1731&331.8&18.034&0.027&17.443&0.012&0.021 \\
265&NGC 5147&18.0&14.293&0.005&13.938&0.002&0.027 \\
266&NGC 5204&3.3&13.167&0.003&13.014&0.002&0.013 \\
267&UGC 8502&149.9&15.943&0.010&15.709&0.005&0.011 \\
268&UGC 8561&107.5&15.819&0.010&15.455&0.005&0.011 \\
269&NGC 5230&105.6&15.235&0.008&14.753&0.003&0.025 \\
270&IRAS 13349+2438&453.5&18.660&0.029&17.344&0.009&0.012 \\
272&UGC 8626&108.8&17.820&0.025&17.193&0.010&0.028 \\
271&NGC 5256&125.2&16.806&0.016&16.292&0.009&0.013 \\
273&NGC 5263&77.5&16.838&0.016&16.394&0.007&0.013 \\
274&MCG 1-35-028&105.1&18.506&0.034&17.683&0.014&0.030 \\
275&IC 910&120.3&18.354&0.031&17.684&0.013&0.017 \\
276&Mrk 268&173.7&19.024&0.042&18.390&0.018&0.016 \\
277&NGC 5278&114.4&16.357&0.013&15.864&0.006&0.009 \\
278&NGC 5273&16.5&17.496&0.022&16.620&0.009&0.010 \\
279&UGC 8685&152.6&16.693&0.015&16.302&0.007&0.014 \\
280&UGC 8686&105.4&17.119&0.018&16.884&0.009&0.026 \\
281&UGC 8696&163.0&18.377&0.032&17.562&0.012&0.008 \\
283&Mrk 796&98.5&18.253&0.023&17.494&0.010&0.026 \\
282&NGC 5297&30.9&14.892&0.007&14.468&0.003&0.014 \\
284&IRAS 13446+1121&104.6&20.637&0.085&19.222&0.029&0.035 \\
285&NGC 5303&23.0&15.486&0.009&15.092&0.004&0.014 \\
286&NGC 5313&30.9&15.799&0.010&15.240&0.004&0.008 \\
287&MCG 3-35-034&178.6&19.052&0.043&18.283&0.018&0.023 \\
288&NGC 5347&39.0&16.295&0.012&15.883&0.006&0.021 \\
289&NGC 5350&30.9&14.834&0.006&14.541&0.003&0.011 \\
291&UGC 8827&85.4&17.459&0.021&16.715&0.009&0.023 \\
293&UGC 8856&137.9&18.939&0.040&18.398&0.019&0.016 \\
294&NGC 5374&68.9&15.673&0.009&15.214&0.004&0.027 \\
295&UGC 8902&114.4&16.935&0.017&16.423&0.008&0.022 \\
296&NGC 5403&37.0&17.581&0.022&17.015&0.011&0.009 \\
297&MCG 7-29-036&144.6&19.870&0.061&19.449&0.030&0.011 \\
299&MCG 5-33-046&116.4&20.712&0.088&19.157&0.029&0.014 \\
300&NGC 5474&5.6&14.318&0.005&14.191&0.003&0.011 \\
302&MCG 6-31-070&155.8&18.301&0.103&18.040&0.016&0.016 \\
303&CGCG 074-129&76.5&20.453&0.274&19.101&0.027&0.024 \\
305&NGC 5515&114.1&17.681&0.023&16.830&0.009&0.008 \\
304&NGC 5520&30.5&15.876&0.010&15.272&0.004&0.018 \\
306&NGC 5526&27.9&17.716&0.024&16.861&0.009&0.012 \\
307&NGC 5522&72.1&17.442&0.021&16.723&0.009&0.024 \\
308&NGC 5541&115.4&16.548&0.014&15.914&0.006&0.011 \\
309&IC 4395&160.5&17.640&0.023&16.684&0.008&0.017 \\
310&UGC 9165&81.3&18.758&0.038&18.123&0.016&0.017 \\
311&Mrk 1490&116.2&20.862&0.093&19.271&0.026&0.018 \\
312&NGC 5585&5.6&13.318&0.003&13.142&0.002&0.016 \\
313&IC 4408&134.9&18.451&0.033&17.681&0.013&0.021 \\
314&NGC 5584&23.1&14.304&0.005&13.926&0.002&0.039 \\
315&NGC 5633&36.5&15.711&0.010&15.130&0.004&0.017 \\
316&NGC 5660&38.9&14.274&0.005&13.868&0.002&0.021 \\
317&NGC 5656&53.7&15.916&0.011&15.337&0.005&0.015 \\
318&NGC 5657&64.4&17.256&0.019&16.935&0.010&0.018 \\
319&CGCG 133-083&190.6&19.039&0.042&18.354&0.017&0.048 \\
320&MCG 7-30-028&116.1&17.796&0.024&17.330&0.011&0.012 \\
321&MCG 6-32-070&127.1&17.141&0.018&16.332&0.007&0.010 \\
322&UGC 9412&138.7&14.825&0.006&14.778&0.003&0.007 \\
324&NGC 5691&19.8&15.272&0.008&14.797&0.004&0.037 \\
325&MCG 9-24-035&137.4&19.059&0.043&17.988&0.017&0.022 \\
326&MCG 9-24-038&166.6&19.254&0.046&18.422&0.018&0.016 \\
327&UGC 9560&23.0&15.360&0.008&15.323&0.004&0.012 \\
328&IC 1076&92.6&16.879&0.016&16.289&0.007&0.031 \\
329&IRAS 14538+1730&432.9&21.043&0.101&19.689&0.032&0.027 \\
330&NGC 5795&38.2&17.965&0.026&17.229&0.011&0.019 \\
331&UGC 9618&145.8&19.178&0.045&18.582&0.019&0.042 \\
332&UGC 9639&157.7&18.082&0.028&17.166&0.011&0.018 \\
333&MCG 6-33-022&194.5&18.685&0.036&18.114&0.017&0.016 \\
334&NGC 5879&15.5&14.598&0.006&14.289&0.003&0.012 \\
335&MCG 9-25-036&160.1&20.680&0.073&19.776&0.029&0.018 \\
336&NGC 5899&43.5&16.323&0.013&15.679&0.006&0.034 \\
337&NGC 5905&58.7&15.358&0.008&15.155&0.005&0.015 \\
338&Mrk 848&173.9&17.811&0.024&17.115&0.010&0.026 \\
339&IC 4553&83.5&17.962&0.026&17.175&0.013&0.051 \\
340&UGC 9922&86.7&16.585&0.014&16.325&0.007&0.017 \\
341&IC 4567&88.6&16.502&0.014&15.888&0.006&0.029 \\
342&MCG 4-37-016&102.9&17.952&0.026&17.238&0.011&0.047 \\
343&NGC 5975&69.3&19.005&0.042&18.201&0.020&0.063 \\
344&NGC 5980&65.2&16.422&0.013&15.844&0.006&0.035 \\
345&NGC 5992&140.2&16.574&0.014&16.112&0.006&0.020 \\
346&NGC 5996&54.0&14.836&0.006&14.475&0.003&0.034 \\
347&IRAS 15519+3537&354.1&20.629&0.062&20.026&0.028&0.025 \\
348&UGC 10099&152.2&16.371&0.013&16.034&0.006&0.018 \\
349&MCG 5-38-006&69.6&18.674&0.036&17.973&0.015&0.051 \\
351&NGC 6027A&70.6&20.475&0.081&19.600&0.031&0.055 \\
350&UGC 10120&138.9&16.570&0.014&16.153&0.007&0.025 \\
352&NGC 6040&177.0&17.108&0.018&16.715&0.010&0.044 \\
353&UGC 10200&31.2&15.086&0.007&14.906&0.004&0.010 \\
354&IRAS 16052+5334&366.1&22.846&0.154&21.821&0.060&0.012 \\
355&IRAS 16053+1836&161.4&20.857&0.093&19.589&0.036&0.043 \\
356&NGC 6090&131.2&16.472&0.013&15.871&0.006&0.020 \\
357&UGC 10273&111.3&17.265&0.019&16.978&0.009&0.051 \\
358&IRAS 16150+2233&278.1&21.368&0.092&20.510&0.043&0.112 \\
359&UGC 10322&69.1&17.798&0.024&17.457&0.013&0.090 \\
360&NGC 6120&134.9&17.360&0.020&16.749&0.009&0.018 \\
361&MCG 3-42-004&171.9&18.171&0.029&17.619&0.014&0.066 \\
362&UGC 10407&124.7&15.789&0.010&15.471&0.005&0.008 \\
363&IRAS 16320+3922&139.4&17.012&0.017&16.610&0.008&0.010 \\
364&NGC 6186&162.8&16.912&0.016&16.223&0.008&0.047 \\
366&UGC 10514&100.5&17.248&0.019&16.882&0.009&0.051 \\
367&IRAS 16435+2154&142.3&20.367&0.076&19.604&0.036&0.047 \\
368&IC 4623&138.5&19.026&0.042&18.568&0.021&0.064 \\
369&IRAS 16516+3030&306.1&21.397&0.085&20.554&0.037&0.034 \\
\hline
\multicolumn{8}{l}
{$^a$Ashby et al.\ 2011; distances based on $H_0 =
  73$\,\kmsmpc}\\
\multicolumn{8}{l}
{$^b$AB magnitude}\\
\multicolumn{8}{l}
{$^c$Milky Way colour excess in magnitudes from \citet{schlegel98}}
\end{longtable}
\end{center}

\clearpage
\begin{center}
\begin{longtable}{rrrrrrrr}
{\footnotesize}
\null\\
\caption[]{SFR measures$^a$ for SFRS galaxies} \\
\hline\hline
SFRS$^b$ & \sfrfuv & \sfrnuv & \sfrr & \sfrrd &
 \sfrfir & \sfrp & \sfrt$^c$ \\
\hline\endhead
\hline\endfoot
1&$-$0.18&0.06&0.97&0.87&0.57&0.74&0.64 \\
2&0.16&0.36&0.89&0.81&0.65&0.84&0.78 \\
3&$-$0.25&$-$0.19&$-$0.63&$-$0.33&$-$0.69&$-$0.53&$-$0.11 \\
5&0.29&0.61&1.53&1.29&1.14&1.22&1.20 \\
8&0.63&0.85&1.30&1.11&0.92&1.31&1.10 \\
9&0.37&0.49&0.70&0.67&0.45&0.45&0.71 \\
10&0.28&0.44&0.69&0.66&0.48&0.68&0.69 \\
11&$-$0.49&$-$0.30&$-$0.22&$-$0.03&$-$0.19&0.03&$-$0.01 \\
12&$-$0.62&$-$0.54&$-$0.78&$-$0.44&$-$0.66&$-$0.77&$-$0.34 \\
13&$-$0.72&$-$0.64&$-$1.58&$-$1.04&$-$1.30&$-$1.77&$-$0.61 \\
14&$-$0.76&$-$0.31&0.84&0.77&0.85&0.66&0.86 \\
15&$-$0.20&$-$0.04&0.19&0.29&0.20&0.48&0.35 \\
16&$-$1.29&$-$1.08&0.39&0.44&0.16&0.35&0.18 \\
17&$-$0.79&$-$0.73&$-$2.44&$-$1.68&$-$1.56&$-$1.58&$-$0.72 \\
18&$-$0.05&0.11&0.43&0.46&0.50&0.36&0.60 \\
19&0.01&0.29&1.67&1.39&1.61&1.14&1.62 \\
20&$-$1.09&$-$0.70&0.92&0.83&0.74&0.71&0.75 \\
21&0.09&0.29&$-$0.06&0.10&$-$0.24&$-$0.09&0.26 \\
22&$-$0.31&$-$0.08&0.15&0.25&0.04&0.31&0.20 \\
24&$-$0.21&0.05&1.65&1.38&1.63&1.02&1.63 \\
25&$-$1.24&$-$0.74&1.00&0.90&1.01&0.81&1.02 \\
29&0.43&0.67&1.60&1.35&1.24&1.48&1.30 \\
31&$-$0.87&$-$0.38&1.34&1.15&1.28&0.86&1.28 \\
32&0.09&0.22&2.36&1.91&1.65&1.43&1.66 \\
33&0.28&0.47&0.84&0.77&0.60&0.78&0.77 \\
34&$-$0.26&$-$0.14&$-$0.08&0.08&$-$0.14&0.21&0.10 \\
35&$-$0.13&0.06&0.74&0.70&0.13&$-$0.39&0.32 \\
36&$-$0.56&$-$0.38&1.26&1.09&2.00&2.27&2.00 \\
37&0.19&0.50&2.11&1.72&1.80&1.37&1.81 \\
38&$-$0.51&$-$0.30&0.32&0.38&0.05&0.23&0.16 \\
39&0.17&0.31&0.34&0.40&0.00&0.33&0.39 \\
40&0.13&0.37&2.51&2.02&1.76&1.23&1.77 \\
41&0.10&0.24&0.38&0.43&0.23&0.45&0.47 \\
42&$-$0.09&0.12&1.21&1.05&0.78&0.80&0.83 \\
43&$-$0.35&0.01&1.49&1.26&1.29&1.35&1.30 \\
44&$-$0.53&$-$0.25&1.01&0.90&0.75&0.96&0.78 \\
45&0.19&0.36&0.45&0.48&0.21&0.60&0.50 \\
46&$-$0.45&$-$0.02&1.05&0.93&0.82&1.01&0.85 \\
48&$-$1.49&$-$0.98&0.06&0.18&$-$0.23&$-$0.42&$-$0.20 \\
49&$-$0.76&$-$0.48&1.30&1.12&1.10&1.19&1.11 \\
50&$-$0.12&0.07&1.01&0.90&0.63&0.58&0.70 \\
51&$-$0.62&$-$0.51&0.95&0.85&0.81&0.92&0.83 \\
52&$-$1.00&$-$0.79&$-$0.08&0.08&$-$0.45&$-$0.10&$-$0.34 \\
53&$-$0.57&$-$0.24&0.82&0.75&0.58&0.69&0.61 \\
54&0.41&0.59&0.99&0.89&0.78&0.92&0.94 \\
55&$-$0.82&$-$0.60&$-$0.52&$-$0.25&$-$0.38&$-$0.30&$-$0.24 \\
56&$-$0.73&$-$0.63&0.55&0.56&0.31&0.58&0.35 \\
57&$-$0.39&$-$0.10&1.13&0.99&1.02&1.03&1.03 \\
58&0.20&0.35&1.01&0.90&0.65&0.72&0.78 \\
59&0.42&0.60&1.46&1.23&1.17&1.25&1.24 \\
60&$-$0.26&$-$0.02&1.47&1.25&1.31&1.33&1.32 \\
61&$-$1.19&$-$0.87&0.95&0.86&0.65&0.77&0.66 \\
62&0.29&0.47&1.32&1.13&0.87&1.07&0.97 \\
63&0.35&0.56&0.99&0.88&0.81&1.15&0.94 \\
64&$-$0.13&0.12&1.42&1.21&1.25&1.33&1.27 \\
65&$-$0.49&$-$0.30&$-$0.11&0.06&$-$0.36&$-$0.14&$-$0.12 \\
67&0.01&0.29&1.23&1.07&0.91&1.15&0.96 \\
68&$-$0.50&$-$0.27&0.80&0.74&0.47&0.71&0.51 \\
69&$-$0.31&$-$0.19&0.02&0.15&$-$0.70&$-$0.82&$-$0.16 \\
70&$-$0.70&$-$0.56&$-$0.67&$-$0.36&$-$0.58&$-$0.44&$-$0.33 \\
71&$-$0.27&$-$0.10&0.21&0.30&$-$0.08&0.18&0.13 \\
72&$-$0.85&$-$0.72&0.07&0.20&$-$0.12&0.18&$-$0.05 \\
73&$-$1.27&$-$1.03&$-$0.04&0.11&$-$0.07&0.04&$-$0.04 \\
74&$-$0.53&$-$0.12&$-$0.11&0.06&$-$0.31&$-$0.11&$-$0.10 \\
75&$-$0.96&$-$0.44&0.53&0.54&0.48&0.56&0.50 \\
76&$-$0.28&0.11&1.22&1.05&0.95&1.06&0.97 \\
77&$-$0.77&$-$0.30&1.42&1.21&1.14&1.29&1.15 \\
78&$-$0.03&0.10&$-$0.05&0.10&$-$0.17&0.21&0.21 \\
79&$-$0.85&$-$0.34&2.26&1.84&1.88&1.20&1.88 \\
80&$-$1.24&$-$0.75&$-$0.61&$-$0.31&$-$0.05&0.18&$-$0.02 \\
81&0.25&0.37&0.84&0.77&0.48&0.61&0.68 \\
82&0.64&0.80&1.21&1.05&1.01&1.17&1.16 \\
83&$-$0.05&0.07&$-$1.42&$-$0.93&$-$0.64&$-$0.78&0.05 \\
84&0.14&0.53&1.84&1.52&1.42&1.54&1.44 \\
85&0.32&0.50&1.24&1.07&0.60&0.91&0.79 \\
86&$-$1.54&$-$0.98&$-$0.68&$-$0.37&$-$0.60&$-$0.63&$-$0.55 \\
88&$-$1.02&$-$0.74&0.57&0.57&0.35&0.58&0.37 \\
89&0.55&0.65&1.48&1.25&1.41&1.32&1.47 \\
90&$-$1.31&$-$1.12&$-$0.47&$-$0.21&$-$0.52&$-$0.41&$-$0.46 \\
91&$-$0.49&$-$0.35&$-$0.20&0.00&0.34&0.65&0.40 \\
92&$-$0.96&$-$0.88&$-$1.52&$-$1.00&$-$1.51&$-$1.78&$-$0.85 \\
93&$-$0.55&$-$0.45&$-$0.33&$-$0.10&$-$0.30&$-$0.33&$-$0.11 \\
95&$-$0.32&$-$0.17&0.55&0.56&0.28&0.50&0.38 \\
96&0.29&0.44&0.78&0.72&0.50&0.65&0.71 \\
97&$-$0.53&$-$0.22&1.18&1.03&0.88&1.02&0.90 \\
98&$-$0.12&0.06&0.14&0.25&$-$0.21&0.24&0.14 \\
99&$-$0.68&$-$0.60&$-$0.55&$-$0.27&$-$0.45&$-$0.58&$-$0.25 \\
100&$-$0.52&$-$0.20&0.82&0.75&0.62&0.67&0.65 \\
101&$-$0.36&$-$0.20&$-$0.11&0.06&$-$0.26&0.08&$-$0.01 \\
102&$-$0.32&$-$0.17&$-$0.45&$-$0.20&$-$0.46&$-$0.15&$-$0.08 \\
103&$-$0.71&$-$0.23&1.00&0.90&0.95&0.93&0.96 \\
104&$-$0.86&$-$0.71&$-$1.02&$-$0.63&$-$1.07&$-$0.86&$-$0.65 \\
105&0.36&0.58&0.91&0.83&0.71&1.04&0.87 \\
106&$-$2.23&$-$1.92&0.27&0.34&$-$0.05&$-$0.79&$-$0.05 \\
107&$-$0.08&0.09&$-$0.31&$-$0.09&$-$0.52&0.65&0.06 \\
109&$-$1.27&$-$0.93&$-$0.05&0.11&0.14&$-$0.03&0.16 \\
110&$-$0.64&$-$0.42&$-$0.25&$-$0.05&$-$0.43&$-$0.15&$-$0.22 \\
111&0.31&0.52&0.94&0.85&0.79&0.96&0.91 \\
112&0.29&0.45&1.77&1.47&1.62&1.36&1.64 \\
113&0.16&0.40&1.01&0.90&0.74&1.00&0.84 \\
114&$-$1.22&$-$0.72&1.15&1.01&0.92&0.80&0.92 \\
115&0.13&0.15&1.17&1.02&1.77&1.02&1.78 \\
116&$-$1.34&$-$0.96&$-$0.30&$-$0.09&$-$0.20&$-$0.06&$-$0.17 \\
117&$-$1.84&$-$1.23&1.10&0.96&1.06&0.65&1.06 \\
118&0.16&0.45&1.69&1.41&0.99&1.27&1.05 \\
119&$-$0.48&$-$0.21&1.25&1.08&1.13&1.26&1.14 \\
120&0.47&0.72&2.16&1.76&1.58&1.42&1.61 \\
121&0.04&0.18&2.36&1.91&1.92&1.83&1.93 \\
122&$-$1.45&$-$1.12&0.09&0.21&$-$0.08&0.20&$-$0.06 \\
124&$-$0.88&$-$0.57&0.28&0.35&0.04&0.12&0.09 \\
126&$-$0.03&0.07&$-$0.32&$-$0.10&$-$0.68&$-$0.80&0.06 \\
127&$-$0.84&$-$0.64&$-$0.50&$-$0.23&$-$0.52&$-$0.29&$-$0.35 \\
128&$-$1.33&$-$1.21&$-$0.17&0.02&$-$0.60&$-$0.57&$-$0.53 \\
129&$-$0.21&0.00&$-$0.31&$-$0.09&$-$0.21&0.23&0.09 \\
130&0.45&0.62&1.13&0.99&0.72&0.84&0.91 \\
131&$-$0.06&0.13&2.12&1.73&1.87&1.43&1.87 \\
132&$-$0.42&$-$0.02&1.31&1.13&1.04&1.23&1.05 \\
133&$-$0.10&0.05&2.07&1.70&2.18&1.53&2.19 \\
134&$-$0.59&$-$0.40&$-$0.41&$-$0.17&$-$0.43&$-$0.03&$-$0.20 \\
135&$-$0.94&$-$0.65&0.42&0.46&0.54&0.45&0.55 \\
136&0.07&0.47&0.80&0.74&0.85&$-$1.01&0.92 \\
137&$-$0.91&$-$0.69&$-$0.36&$-$0.13&$-$0.52&$-$0.23&$-$0.37 \\
139&0.08&0.42&1.09&0.96&0.87&1.22&0.93 \\
140&0.12&0.34&1.06&0.94&0.76&0.91&0.85 \\
142&$-$0.68&$-$0.52&$-$0.39&$-$0.15&$-$0.57&$-$0.29&$-$0.32 \\
143&$-$0.74&$-$0.64&$-$0.96&$-$0.58&$-$0.97&$-$0.96&$-$0.54 \\
144&$-$1.20&$-$0.55&1.58&1.33&1.32&0.91&1.32 \\
145&0.82&1.01&1.49&1.26&1.06&1.29&1.25 \\
146&$-$0.37&$-$0.08&1.32&1.13&1.02&0.95&1.04 \\
148&$-$0.19&0.21&1.33&1.14&0.98&1.31&1.01 \\
149&0.49&0.65&1.22&1.06&0.87&0.71&1.02 \\
150&$-$0.21&$-$0.06&0.85&0.78&0.70&0.94&0.75 \\
152&$-$1.40&$-$0.63&0.01&0.15&0.05&0.21&0.07 \\
153&$-$0.67&$-$0.37&1.14&1.00&0.93&1.06&0.94 \\
155&$-$0.90&$-$0.64&$-$0.12&0.05&$-$0.27&$-$0.10&$-$0.18 \\
156&$-$0.90&$-$0.73&$-$0.61&$-$0.31&$-$0.61&$-$0.36&$-$0.43 \\
157&$-$1.27&$-$0.90&0.88&0.80&0.40&0.49&0.41 \\
158&$-$0.51&$-$0.11&1.06&0.93&0.80&1.12&0.82 \\
159&$-$0.52&$-$0.03&1.14&1.00&0.92&1.20&0.94 \\
160&$-$0.56&$-$0.10&1.91&1.58&1.52&1.49&1.52 \\
161&0.57&0.62&0.88&0.80&0.39&0.68&0.79 \\
162&0.12&0.26&0.84&0.77&0.58&0.63&0.71 \\
163&$-$0.20&0.06&0.74&0.70&0.54&0.90&0.61 \\
164&$-$1.07&$-$0.86&$-$0.38&$-$0.14&$-$0.59&$-$0.31&$-$0.47 \\
165&$-$0.60&$-$0.29&1.05&0.93&0.68&0.85&0.70 \\
166&$-$0.98&$-$0.85&$-$1.01&$-$0.61&$-$1.15&$-$0.93&$-$0.75 \\
167&$-$0.88&$-$0.53&0.67&0.65&1.14&0.84&1.14 \\
168&$-$0.31&$-$0.17&0.88&0.81&0.56&0.77&0.62 \\
169&$-$0.54&$-$0.12&1.01&0.90&0.79&0.94&0.81 \\
170&$-$1.55&$-$1.39&$-$1.48&$-$0.97&$-$1.37&$-$1.13&$-$1.15 \\
171&$-$0.63&$-$0.40&$-$0.86&$-$0.51&$-$0.50&$-$0.02&$-$0.26 \\
172&$-$2.05&$-$1.59&$-$1.32&$-$0.85&$-$1.12&$-$0.92&$-$1.07 \\
173&$-$0.67&$-$0.19&1.13&0.99&1.07&1.14&1.08 \\
174&$-$0.39&$-$0.31&$-$0.50&$-$0.24&$-$0.73&$-$0.49&$-$0.23 \\
175&$-$0.30&$-$0.18&$-$0.56&$-$0.28&$-$0.70&$-$0.47&$-$0.15 \\
176&$-$1.92&$-$1.36&$-$1.58&$-$1.04&$-$1.13&$-$1.08&$-$1.06 \\
177&$-$3.25&$-$3.01&$-$1.12&$-$0.70&$-$1.22&$-$1.31&$-$1.22 \\
178&$-$0.01&0.23&0.64&0.62&0.21&0.52&0.42 \\
179&$-$0.26&$-$0.11&$-$0.22&$-$0.02&$-$0.41&$-$0.17&$-$0.03 \\
181&$-$0.73&$-$0.52&$-$0.45&$-$0.20&$-$0.47&$-$0.13&$-$0.28 \\
183&$-$1.12&$-$1.03&$-$2.55&$-$1.77&$-$1.49&$-$1.50&$-$0.97 \\
184&$-$1.57&$-$1.30&$-$0.43&$-$0.18&$-$0.52&$-$0.25&$-$0.48 \\
185&$-$0.60&$-$0.25&0.44&0.47&0.56&0.61&0.59 \\
186&$-$0.32&$-$0.15&$-$0.51&$-$0.24&$-$0.29&$-$0.05&0.00 \\
187&$-$1.11&$-$0.74&$-$0.92&$-$0.55&$-$0.49&$-$0.17&$-$0.40 \\
188&$-$1.05&$-$0.85&$-$1.44&$-$0.94&$-$1.99&$-$1.48&$-$1.00 \\
189&$-$0.51&$-$0.21&1.05&0.93&0.64&0.57&0.67 \\
190&$-$2.27&$-$1.92&$-$0.51&$-$0.24&$-$0.82&$-$0.67&$-$0.81 \\
191&$-$0.32&$-$0.12&0.26&0.34&0.25&0.36&0.35 \\
192&$-$0.42&$-$0.27&$-$0.38&$-$0.14&$-$0.60&$-$0.51&$-$0.20 \\
193&$-$0.99&$-$0.64&$-$0.61&$-$0.32&$-$0.53&$-$0.54&$-$0.40 \\
194&$-$0.38&$-$0.24&$-$0.11&0.06&0.03&0.09&0.17 \\
196&$-$0.68&$-$0.54&$-$0.39&$-$0.15&$-$0.85&$-$0.60&$-$0.46 \\
197&$-$0.28&$-$0.12&$-$0.02&0.13&$-$0.05&0.14&0.15 \\
199&$-$0.61&$-$0.41&$-$1.06&$-$0.65&$-$0.52&$-$0.22&$-$0.26 \\
200&$-$2.05&$-$1.71&$-$2.00&$-$1.36&$-$1.49&$-$1.31&$-$1.38 \\
201&$-$2.02&$-$1.33&$-$1.53&$-$1.01&$-$0.86&$-$0.84&$-$0.83 \\
202&$-$0.96&$-$0.72&0.10&0.22&$-$0.41&$-$0.34&$-$0.30 \\
203&$-$1.18&$-$0.89&$-$1.02&$-$0.62&$-$0.77&$-$0.50&$-$0.63 \\
205&$-$0.73&$-$0.58&$-$0.51&$-$0.24&$-$0.74&$-$0.50&$-$0.43 \\
207&$-$1.61&$-$1.34&$-$1.05&$-$0.64&$-$0.70&&$-$0.65 \\
209&$-$0.25&$-$0.01&1.07&0.94&0.77&0.98&0.81 \\
210&$-$0.62&$-$0.39&$-$0.83&$-$0.48&$-$0.89&&$-$0.44 \\
211&$-$1.20&$-$1.13&$-$1.60&$-$1.06&$-$1.50&$-$2.31&$-$1.02 \\
212&$-$0.37&$-$0.23&$-$0.29&$-$0.08&$-$0.49&$-$0.33&$-$0.13 \\
213&$-$0.61&$-$0.40&$-$1.03&$-$0.63&$-$0.49&$-$0.04&$-$0.24 \\
214&0.06&0.27&1.44&1.22&1.14&1.19&1.18 \\
216&$-$0.67&$-$0.54&$-$1.23&$-$0.78&$-$1.05&$-$1.05&$-$0.52 \\
217&$-$1.78&$-$1.49&$-$0.34&$-$0.11&$-$0.53&$-$0.26&$-$0.51 \\
218&$-$1.29&$-$1.17&$-$1.48&$-$0.96&$-$1.39&$-$1.08&$-$1.03 \\
219&$-$0.87&$-$0.73&$-$0.66&$-$0.36&$-$0.70&$-$0.46&$-$0.48 \\
222&$-$1.48&$-$0.90&0.58&0.58&0.69&0.65&0.70 \\
223&$-$0.20&0.08&1.21&1.05&0.80&0.94&0.84 \\
224&$-$0.53&$-$0.45&$-$0.70&$-$0.39&$-$0.83&$-$0.92&$-$0.35 \\
226&$-$0.64&$-$0.31&1.68&1.40&1.31&1.43&1.32 \\
228&$-$0.57&$-$0.53&$-$1.35&$-$0.87&$-$1.10&$-$1.15&$-$0.46 \\
229&0.28&0.45&1.02&0.91&0.85&0.92&0.95 \\
231&$-$1.68&$-$0.84&1.95&1.61&1.89&1.40&1.89 \\
233&$-$0.78&$-$0.43&1.56&1.31&1.41&1.05&1.41 \\
234&$-$1.40&$-$1.29&$-$0.97&$-$0.59&$-$0.85&$-$0.89&$-$0.74 \\
235&$-$0.09&0.09&1.37&1.17&0.91&1.18&0.95 \\
236&$-$0.72&$-$0.56&$-$0.52&$-$0.25&$-$0.77&$-$0.86&$-$0.45 \\
237&$-$2.16&$-$1.89&$-$0.62&$-$0.32&$-$1.15&$-$1.14&$-$1.11 \\
238&$-$0.35&$-$0.22&$-$1.25&$-$0.80&$-$0.73&$-$0.57&$-$0.20 \\
240&0.47&0.65&1.50&1.27&1.05&1.24&1.15 \\
241&0.27&0.41&1.38&1.18&1.27&1.05&1.31 \\
242&$-$0.45&$-$0.37&$-$0.46&$-$0.20&$-$0.75&$-$1.34&$-$0.27 \\
244&$-$0.75&$-$0.47&1.50&1.27&1.29&1.11&1.29 \\
245&0.57&0.71&1.76&1.46&1.20&1.45&1.29 \\
246&$-$0.14&0.31&1.50&1.27&1.22&1.36&1.24 \\
247&$-$0.55&$-$0.33&0.57&0.57&0.35&0.50&0.40 \\
248&$-$1.26&$-$0.83&1.31&1.12&1.16&0.81&1.16 \\
249&$-$1.31&$-$1.06&$-$0.56&$-$0.28&$-$0.62&$-$0.47&$-$0.54 \\
250&0.06&0.29&0.53&0.54&0.24&0.65&0.46 \\
251&$-$1.20&$-$0.69&0.37&0.42&1.30&1.04&1.30 \\
252&$-$1.76&$-$0.98&0.81&0.75&1.12&$-$0.01&1.12 \\
253&0.50&0.62&1.78&1.48&1.53&2.13&1.57 \\
254&0.08&0.35&1.04&0.92&0.99&1.04&1.04 \\
255&0.34&0.58&1.49&1.26&1.22&1.33&1.28 \\
256&$-$0.55&$-$0.24&1.14&1.00&0.96&0.98&0.98 \\
257&$-$0.16&0.14&1.91&1.57&1.68&1.44&1.69 \\
258&0.00&0.32&2.16&1.77&1.16&1.32&1.18 \\
259&$-$0.43&$-$0.21&1.34&1.14&1.22&1.17&1.23 \\
260&$-$0.76&$-$0.62&$-$0.96&$-$0.58&$-$1.03&$-$1.41&$-$0.57 \\
262&0.63&0.82&1.30&1.12&1.01&1.33&1.16 \\
263&0.57&0.33&2.16&1.76&2.33&3.01&2.34 \\
264&0.74&0.97&1.94&1.60&1.46&1.66&1.53 \\
265&$-$0.28&$-$0.14&$-$0.24&$-$0.04&$-$0.49&$-$0.26&$-$0.07 \\
266&$-$1.36&$-$1.30&$-$1.85&$-$1.25&$-$2.04&$-$2.19&$-$1.28 \\
267&0.85&0.94&1.09&0.96&1.04&1.01&1.25 \\
268&0.61&0.75&1.38&1.18&1.06&1.26&1.19 \\
269&0.88&1.06&1.06&0.94&0.84&1.20&1.16 \\
270&0.73&1.25&2.45&1.98&1.60&2.41&1.66 \\
271&0.35&0.56&2.15&1.75&1.53&1.39&1.56 \\
272&$-$0.12&0.12&0.72&0.68&0.68&0.76&0.75 \\
273&$-$0.08&0.10&1.21&1.05&0.85&1.02&0.90 \\
274&$-$0.42&$-$0.09&1.09&0.96&0.93&0.89&0.95 \\
275&$-$0.29&$-$0.02&1.60&1.34&1.29&1.03&1.30 \\
276&$-$0.24&0.01&1.96&1.61&1.06&1.03&1.08 \\
277&0.44&0.64&1.33&1.14&0.88&0.99&1.01 \\
278&$-$1.69&$-$1.34&$-$1.17&$-$0.74&$-$1.21&$-$1.20&$-$1.09 \\
279&0.58&0.73&1.40&1.19&1.26&1.22&1.34 \\
280&0.12&0.21&1.06&0.94&0.86&0.95&0.93 \\
281&$-$0.06&0.26&2.47&2.00&2.19&1.62&2.19 \\
282&$-$0.09&0.08&0.19&0.28&$-$0.04&0.32&0.24 \\
283&$-$0.39&$-$0.09&2.12&1.73&0.96&0.93&0.98 \\
284&$-$1.26&$-$0.70&1.20&1.04&0.98&1.00&0.98 \\
285&$-$0.59&$-$0.43&$-$0.03&0.12&$-$0.27&$-$0.02&$-$0.10 \\
286&$-$0.47&$-$0.25&0.48&0.50&0.09&0.46&0.20 \\
287&$-$0.20&0.10&1.67&1.39&1.12&1.22&1.14 \\
288&$-$0.43&$-$0.27&$-$0.22&$-$0.02&$-$0.17&0.10&0.02 \\
289&$-$0.08&0.04&0.05&0.18&0.00&0.38&0.27 \\
291&$-$0.21&0.09&1.07&0.95&0.91&0.90&0.94 \\
293&$-$0.41&$-$0.19&1.46&1.24&1.10&1.16&1.12 \\
294&0.34&0.52&0.91&0.82&0.67&0.93&0.84 \\
295&0.25&0.46&1.06&0.94&0.77&0.98&0.89 \\
296&$-$1.03&$-$0.80&0.56&0.56&0.24&0.45&0.26 \\
297&$-$0.75&$-$0.59&0.88&0.81&1.19&0.89&1.19 \\
299&$-$1.27&$-$0.65&1.00&0.89&1.02&0.86&1.02 \\
300&$-$1.35&$-$1.30&$-$1.54&$-$1.01&$-$1.61&$-$1.88&$-$1.16 \\
302&$-$0.04&0.06&1.49&1.26&1.29&1.27&1.31 \\
303&$-$1.50&$-$0.96&0.50&0.51&0.71&0.60&0.71 \\
304&$-$0.48&$-$0.24&$-$0.01&0.13&$-$0.12&0.15&0.04 \\
305&$-$0.09&0.25&1.42&1.21&0.93&1.19&0.97 \\
306&$-$1.32&$-$0.98&$-$0.06&0.10&$-$0.24&$-$0.09&$-$0.20 \\
307&$-$0.34&$-$0.06&0.81&0.75&0.54&0.68&0.59 \\
308&0.38&0.63&1.49&1.26&1.03&1.20&1.12 \\
309&0.25&0.63&1.61&1.35&1.37&1.35&1.40 \\
310&$-$0.79&$-$0.54&1.16&1.01&0.85&1.04&0.86 \\
311&$-$1.32&$-$0.68&1.28&1.11&1.36&1.10&1.36 \\
312&$-$0.94&$-$0.87&$-$1.38&$-$0.89&$-$1.71&$-$1.73&$-$0.87 \\
313&$-$0.21&0.09&1.41&1.20&1.08&1.29&1.10 \\
314&$-$0.03&0.12&$-$0.14&0.04&$-$0.32&$-$0.06&0.15 \\
315&$-$0.26&$-$0.04&0.42&0.46&0.14&0.49&0.28 \\
316&0.38&0.54&0.55&0.56&0.32&0.58&0.65 \\
317&$-$0.02&0.21&0.63&0.61&0.48&0.77&0.60 \\
318&$-$0.39&$-$0.26&0.48&0.50&0.28&0.51&0.37 \\
319&$-$0.06&0.21&1.45&1.23&1.40&1.35&1.41 \\
320&$-$0.11&0.07&1.08&0.95&0.95&1.03&0.99 \\
321&0.22&0.54&1.40&1.19&1.17&1.39&1.22 \\
322&1.22&1.23&1.18&1.03&1.03&1.77&1.44 \\
324&$-$0.55&$-$0.37&$-$0.28&$-$0.07&$-$0.37&$-$0.22&$-$0.15 \\
325&$-$0.44&$-$0.01&1.19&1.03&1.20&0.97&1.21 \\
326&$-$0.37&$-$0.04&1.43&1.22&1.11&1.33&1.12 \\
327&$-$0.54&$-$0.53&$-$0.66&$-$0.35&$-$0.92&$-$1.53&$-$0.39 \\
328&0.12&0.36&1.10&0.97&0.67&0.89&0.78 \\
329&$-$0.22&0.32&2.21&1.80&1.92&1.74&1.93 \\
330&$-$1.12&$-$0.83&0.28&0.35&0.15&0.38&0.17 \\
331&$-$0.37&$-$0.13&2.15&1.75&1.68&1.67&1.68 \\
332&0.06&0.42&1.80&1.49&1.30&1.53&1.33 \\
333&0.00&0.22&2.55&2.06&1.50&1.54&1.51 \\
334&$-$0.58&$-$0.46&$-$0.48&$-$0.22&$-$0.53&$-$0.16&$-$0.25 \\
335&$-$0.97&$-$0.61&0.96&0.86&0.81&0.89&0.81 \\
336&$-$0.30&$-$0.05&0.75&0.70&0.42&0.79&0.49 \\
337&0.28&0.36&0.69&0.66&0.53&1.17&0.72 \\
338&0.28&0.56&2.03&1.67&1.86&1.59&1.87 \\
339&$-$0.33&$-$0.02&2.21&1.80&2.27&1.30&2.28 \\
340&0.14&0.24&1.07&0.94&0.73&0.79&0.83 \\
341&0.23&0.47&1.12&0.98&0.81&1.11&0.92 \\
342&$-$0.16&0.12&1.18&1.02&0.81&0.59&0.85 \\
343&$-$0.87&$-$0.56&0.87&0.80&0.76&0.77&0.77 \\
344&0.01&0.24&0.88&0.80&0.71&1.00&0.79 \\
345&0.57&0.75&1.35&1.15&1.05&1.26&1.17 \\
346&0.48&0.62&0.78&0.73&0.57&0.71&0.83 \\
347&$-$0.23&0.01&2.00&1.64&1.57&1.62&1.58 \\
348&0.71&0.85&1.38&1.18&1.12&1.18&1.27 \\
349&$-$0.78&$-$0.50&0.57&0.57&0.57&0.61&0.59 \\
350&0.58&0.74&0.87&0.80&0.60&1.00&0.89 \\
351&$-$1.47&$-$1.13&0.60&0.59&0.23&0.11&0.24 \\
352&0.64&0.79&2.33&1.89&0.86&1.00&1.06 \\
353&$-$0.17&$-$0.10&$-$0.11&0.06&$-$0.42&$-$0.58&0.02 \\
354&$-$1.13&$-$0.73&1.98&1.63&1.62&1.71&1.62 \\
355&$-$0.95&$-$0.44&1.41&1.20&1.20&1.29&1.21 \\
356&0.55&0.79&1.77&1.47&1.51&1.51&1.55 \\
357&0.19&0.31&1.29&1.11&0.89&0.76&0.97 \\
358&$-$0.45&$-$0.11&1.33&1.14&1.50&1.10&1.50 \\
359&$-$0.30&$-$0.17&0.91&0.83&0.55&0.80&0.61 \\
360&0.21&0.46&1.64&1.37&1.37&1.52&1.40 \\
361&0.26&0.48&1.33&1.14&1.12&1.20&1.18 \\
362&0.74&0.87&1.31&1.12&0.92&0.96&1.14 \\
363&0.36&0.51&0.95&0.85&0.70&0.80&0.86 \\
364&0.65&0.92&1.51&1.28&1.41&1.58&1.48 \\
366&0.11&0.26&1.36&1.16&0.90&0.97&0.97 \\
367&$-$0.85&$-$0.54&1.22&1.06&1.11&1.02&1.12 \\
368&$-$0.28&$-$0.10&1.17&1.02&1.01&1.07&1.04 \\
369&$-$0.64&$-$0.30&1.94&1.60&1.69&1.69&1.69 \\
\hline
\multicolumn{8}{l}
{$^a$in units of $\rm \log(SFR/\Msol\,yr^{-1})$}\\
\multicolumn{8}{l}
{$^b$Ashby et al. 2011}\\
\multicolumn{8}{l}
{$^c$$\log{\rm [SFR(FIR) + SFR(FUV)]}$}
\end{longtable}
\end{center}

\bsp    
\label{lastpage}
\end{document}